\documentclass[preprint2]{aastex}
\usepackage{color}
\usepackage{graphicx}
\usepackage{bm}
\usepackage{amsmath}
\let\la=\langle
\let\ra=\rangle
\newcommand{\ue}{\mathrm{e}}

\begin{document}
\title{POLARIZED LINE FORMATION IN MULTI-DIMENSIONAL MEDIA. V. 
EFFECTS OF ANGLE-DEPENDENT PARTIAL
FREQUENCY REDISTRIBUTION}
\author{L.~S.~Anusha$^{1}$ and K.~N.~Nagendra$^{1}$}
\affil{$^1$Indian Institute of Astrophysics, Koramangala,
2nd Block, Bangalore 560 034, India}

\begin{abstract}
The solution of polarized radiative transfer equation with angle-dependent (AD)
partial frequency redistribution (PRD) is a challenging problem. Modeling 
the observed, linearly polarized strong resonance lines in the 
solar spectrum often requires the solution of the AD line transfer 
problems in one-dimensional (1D) or multi-dimensional (multi-D) geometries. 
The purpose of this paper is to develop an understanding of the 
relative importance of the AD PRD effects and the multi-D transfer effects 
and particularly their combined influence on the line polarization. 
This would help in a quantitative analysis of the second solar 
spectrum (the linearly polarized spectrum of the Sun). 
We consider both non-magnetic and magnetic media.
In this paper we reduce the Stokes vector transfer equation to a 
simpler form using a Fourier decomposition technique for multi-D media. 
A fast numerical method is also devised to solve the concerned multi-D 
transfer problem. The numerical results are presented for a two-dimensional 
medium with a moderate optical thickness (effectively thin), and are
computed for a collisionless frequency redistribution.
We show that the AD PRD effects are significant, and can not be ignored 
in a quantitative fine analysis of the line polarization. 
These effects are accentuated by the finite dimensionality of the medium 
(multi-D transfer). The presence of magnetic fields (Hanle effect) modifies 
the impact of these two effects to a considerable extent.
\end{abstract}

\keywords{line: formation -- radiative transfer -- polarization --
scattering -- magnetic fields -- Sun: atmosphere}

\section{INTRODUCTION}
\label{introduction}
The solution of polarized line transfer equation with angle-dependent (AD) 
partial frequency redistribution (PRD) has always remained one of the 
difficult areas in the astrophysical line formation theory. The difficulty 
stems from the inextricable coupling between frequency and angle 
variables, which are hard to represent using finite resolution grids. 
Equally challenging is the 
problem of polarized line radiative transfer (RT) equation in 
multi-dimensional (multi-D) media. There existed lack of formulations that 
reduce the complexity of multi-D transfer, when PRD is taken into account. 
In the first three papers of the series on multi-D transfer (see 
Anusha \& Nagendra 2011a - Paper I; Anusha et al 2011a - Paper II; 
Anusha \& Nagendra 2011b - Paper III), 
we formulated and solved the transfer problem using angle-averaged (AA) 
PRD. The Fourier decomposition technique for the AD PRD to solve 
transfer problem in one-dimensional (1D) media including Hanle effect 
was formulated by \citet{hf09}. 
In Anusha \& Nagendra (2011c - hereafter Paper IV), we 
extended this technique to handle multi-D RT with the AD PRD. 
In this paper we apply the technique presented 
in Paper IV to establish several benchmark solutions of the corresponding 
line transfer problem. 
A historical account of the work on polarized RT with the AD 
PRD in 1D planar media, and the related topics is 
given in detail, in Table 1 of Paper IV. Therefore we do not repeat here. 

In Section~\ref{frte} 
we present the multi-D polarized RT equation, 
expressed in terms of irreducible Fourier 
coefficients, denoted by $\tilde{\bm{\mathcal{I}}}^{(k)}$ and 
$\tilde{\bm{\mathcal{S}}}^{(k)}$, where $k$ is the index of the terms in the 
Fourier series expansion of the Stokes vector $\bm {I}$ and the Stokes 
source vector $\bm {S}$. Section~\ref{numerics} describes the numerical 
method of solving the concerned transfer equation.  
Section~\ref{results} is devoted to a discussion of the results. Conclusions 
are presented in Section~\ref{conclusions}. 

\section{POLARIZED TRANSFER EQUATION IN A MULTI-D MEDIUM}
\label{frte}
The multi-D transfer equation written in terms of the Stokes parameters
and the relevant expressions for the Stokes source vectors 
(for line and continuum) in a two-level atom model with unpolarized
ground level, involving the AD PRD matrices is well explained
in Section~2 of Paper IV. 
All these equations can be expressed in terms of `irreducible spherical 
tensors' (see Section~3 of Paper IV). Further,
in Section~4 of Paper IV we developed a decomposition technique 
to simplify this RT equation using Fourier series expansions of the 
AD PRD functions. 
Here we describe a variant of the method presented in Paper IV, 
which is more useful in practical applications involving polarized
RT in magnetized two-dimensional (2D) and three-dimensional (3D) atmospheres.
\subsection{THE RADIATIVE TRANSFER EQUATION IN TERMS OF IRREDUCIBLE SPHERICAL
TENSORS}
Let $\bm{I}=(I,Q,U)^T$ be the Stokes vector 
and $\bm{S}=(S_I,S_Q,S_U)^T$ denote the 
Stokes source vector \citep[see][]{chandra60}.
We introduce vectors $\bm{\mathcal S}$ and $\bm{\mathcal I}$
given by
\begin{eqnarray}
&&\bm{\mathcal S}=(S^0_0,
S^2_0,S^{2,{\rm x}}_1, S^{2,{\rm y}}_1,
S^{2,{\rm x}}_2, S^{2,{\rm y}}_2)^T,\nonumber\\
&&\bm{\mathcal I}=(I^0_0,
I^2_0, I^{2,{\rm x}}_1, I^{2,{\rm y}}_1,
I^{2,{\rm x}}_2, I^{2,{\rm y}}_2)^T.
\end{eqnarray}
These quantities are related to the Stokes parameters 
\citep[see e.g.,][]{hf07} through
\begin{eqnarray}
&&I(\bm{r}, \bm{\Omega}, x) = I^0_0 +
\frac{1}{2 \sqrt{2}} (3 \cos^2\theta -1) I^2_0 \nonumber \\
&&-\sqrt{3} \cos \theta \sin \theta (I^{2, {\rm x}}_1 
\cos \varphi-I^{2, {\rm y}}_1 \sin \varphi) \nonumber \\ 
&&+ \frac{\sqrt{3}}{2} (1-\cos^2\theta)
(I^{2, {\rm x}}_2 \cos 2 \varphi-I^{2, {\rm y}}_2 \sin 2 \varphi), \nonumber \\
\label{transform-1}
\end{eqnarray}

\begin{eqnarray}
&&Q(\bm{r}, \bm{\Omega}, x)= -\frac{3}{2 \sqrt{2}} 
(1- \cos^2\theta) I^2_0 \nonumber \\
&&-\sqrt{3} \cos \theta \sin \theta (I^{2, {\rm x}}_1 
\cos \varphi-I^{2, {\rm y}}_1 \sin \varphi) \nonumber \\ 
&&-\frac{\sqrt{3}}{2} (1+\cos^2\theta)
(I^{2, {\rm x}}_2 \cos 2 \varphi-I^{2, {\rm y}}_2 \sin 2 \varphi),\nonumber \\
\label{transform-2}
\end{eqnarray}
\begin{eqnarray}
&&U(\bm{r}, \bm{\Omega}, x) = \sqrt{3} \sin \theta
(I^{2, {\rm x}}_1 \sin \varphi+I^{2, {\rm y}}_1 \cos \varphi) \nonumber \\ 
&&+ \sqrt{3} \cos \theta 
(I^{2, {\rm x}}_2 \sin 2 \varphi+I^{2, {\rm y}}_2 \cos 2 \varphi).
\label{transform-3}
\end{eqnarray}
We note here that the 
quantities $I^0_0$, $I^2_0$, $I^{2,{\rm x}}_1$, $I^{2,{\rm y}}_1$,
$I^{2,{\rm x}}_2$ and $I^{2,{\rm y}}_2$ also depend on
the variables $\bm{r}$, $\bm{\Omega}$ and $x$ (defined below).

For a given ray defined by the direction $\bm{\Omega}$, the 
vectors $\bm{\mathcal S}$ and $\bm{\mathcal I}$ satisfy the 
RT equation (see Section 3 of paper IV) 
\begin{eqnarray}
&&-\frac{1}{\kappa_{\rm tot}(\bm{r}, x)}\bm{\Omega} \cdot
\bm{\nabla}\bm{\mathcal{I}}(\bm{r}, \bm{\Omega}, x) = \nonumber \\
\!\!\!\!\!\!&&[\bm{\mathcal{I}}(\bm{r}, \bm{\Omega}, x)-
\bm{\mathcal S}(\bm{r}, \bm{\Omega}, x)].
\label{rte-reduced}
\end{eqnarray}
It is useful to note that the above equation was referred to as `irreducible RT
equation' in Paper IV. Indeed, for the AA PRD problems,
the quantities $\bm{\mathcal I}$ and $\bm{\mathcal S}$ are already
in the irreducible form. But for the AD PRD problems,
$\bm{\mathcal I}$ and $\bm{\mathcal S}$ can further be reduced to
$\tilde{\bm{\mathcal I}}^{(k)}$ and $\tilde{\bm{\mathcal S}}^{(k)}$
using Fourier series expansions. 
Here $\bm{r}$ is the position vector of the point
in the medium with coordinates $({\rm x},{\rm y}, {\rm z})$.
The unit vector $\bm{\Omega}=(\eta,\gamma,\mu)=
(\sin\theta\,\cos \varphi\,,\sin\theta\,\sin \varphi\,,\cos \theta)$ 
defines the direction cosines of the ray with respect to
the atmospheric normal (the $Z$-axis), where $\theta$ and 
$\varphi$ are the polar 
and azimuthal angles of the ray.
Total opacity $\kappa_{\rm tot}(\bm{r}, x)$ is given by
\begin{equation}
\kappa_{\rm tot}(\bm{r}, x)=\kappa_l(\bm{r}) \phi(x)+\kappa_c(\bm{r}),
\label{ktot}
\end{equation}
where $\kappa_l$ is the
frequency averaged line opacity,
$\phi$ is the Voigt profile function and $\kappa_c$ is the continuum opacity.
Frequency is measured in reduced units, namely
$x=(\nu-\nu_0)/\Delta \nu_D$ where $\Delta \nu_D$ is the Doppler width.

For a two-level atom model
with unpolarized ground level, $\bm{\mathcal S}(\bm{r}, \bm{\Omega}, x)$
has contributions from the line and the continuum sources. It takes
the form 
\begin{equation}
\bm{\mathcal S}(\bm{r}, \bm{\Omega}, x)=p_x 
\bm{\mathcal S}_{l}(\bm{r}, \bm{\Omega}, x)
+(1-p_x)\bm{\mathcal S}_C(\bm{r}, x),
\label{stot-reduced}
\end{equation}
with
\begin{equation}
p_x=\kappa_l(\bm{r}) \phi(x) / \kappa_{\rm tot}(\bm{r}, x).
\label{px}
\end{equation}
The line source vector is written as
\begin{eqnarray}
\!\!\!\!\!\!&&\bm{\mathcal S}_{l}(\bm{r}, \bm{\Omega}, x)=
\bm{\mathcal G}(\bm{r})+\frac{1}{\phi(x)} 
\int_{-\infty}^{+\infty} dx' \nonumber \\
\!\!\!\!\!\!&& \times
\oint\frac{d \bm{\Omega}'} {4 \pi} 
\hat{W}\Big\{\hat{M}_{\rm II}(\bm{B},x,x')
r_{\rm II}(x, x', \bm{\Omega}, \bm{\Omega'}) \nonumber \\
\!\!\!\!\!\!&&+\hat{M}_{\rm III}(\bm{B},x,x')
r_{\rm III}(x, x',\bm{\Omega}, \bm{\Omega'}) \Big\}
\hat{\Psi}(\bm{\Omega}') \nonumber \\
&&\times \bm{\mathcal{I}}(\bm{r}, \bm{\Omega}',x'),
\label{sl-reduced}
\end{eqnarray}
with $\bm{\mathcal G}(\bm{r})=(\epsilon {B}_{\nu}(\bm{r}),0,0,0,0,0)^T$ and
the unpolarized continuum source vector
$\bm{\mathcal S}_C(\bm{r}, x)$
=$(S_C(\bm{r},x),0,0,0,0,0)^T$.
We assume that $S_C(\bm{r}, x)=B_{\nu}(\bm{r})$ 
with ${B}_{\nu}(\bm{r})$ being the Planck function.
The thermalization parameter $\epsilon=\Gamma_I/(\Gamma_R+\Gamma_I)$
with $\Gamma_I$ and $\Gamma_R$ being the inelastic collision rate and the
radiative de-excitation rate respectively.
The damping parameter is computed using $a=a_R [1+(\Gamma_E+\Gamma_I)/\Gamma_R]$
where $a_R={\Gamma_R}/{4 \pi \Delta \nu_D}$ and $\Gamma_E$ is the
elastic collision rate.
The matrix $\hat{\Psi}$ represents the reduced phase matrix for
the Rayleigh scattering. Its elements are listed in Appendix D 
of Paper III. The elements of the 
matrices $\hat{M}_{\rm II,III}(\bm{B},x,x')$ for the Hanle
effect are derived in \citet{bom97a,bom97b}. The dependence of the matrices
$\hat{M}_{\rm II,III}(\bm{B},x,x')$ on $x$ and $x'$ is related to the
definitions of the frequency domains \citep[see approximation 
level II of][]{bom97b}.
$\hat{W}$ is a diagonal matrix written as
\begin{equation}
\hat{W}=\textrm{diag}\{W_0,W_2,W_2,W_2,W_2,W_2\}.
\label{w}
\end{equation}
Here the weight $W_0=1$ and the weight $W_2$ depends on the line under
consideration \citep[see][]{ll04}. In this paper we take $W_2=1$.
$r_{\rm II,III}$ are the AD PRD
functions of \citet[][]{hum62} which depend explicitly
on the scattering angle $\Theta$, defined through
$\cos \Theta = \bm{\Omega} \cdot \bm{\Omega'}$ computed using
\begin{equation}
\cos \Theta = \mu \mu' + \sqrt{(1-\mu^2)(1-\mu'^2}) \cos (\varphi'-\varphi).
\label{captheta}
\end{equation}

\begin{figure*}
\centering
\includegraphics[scale=0.5]{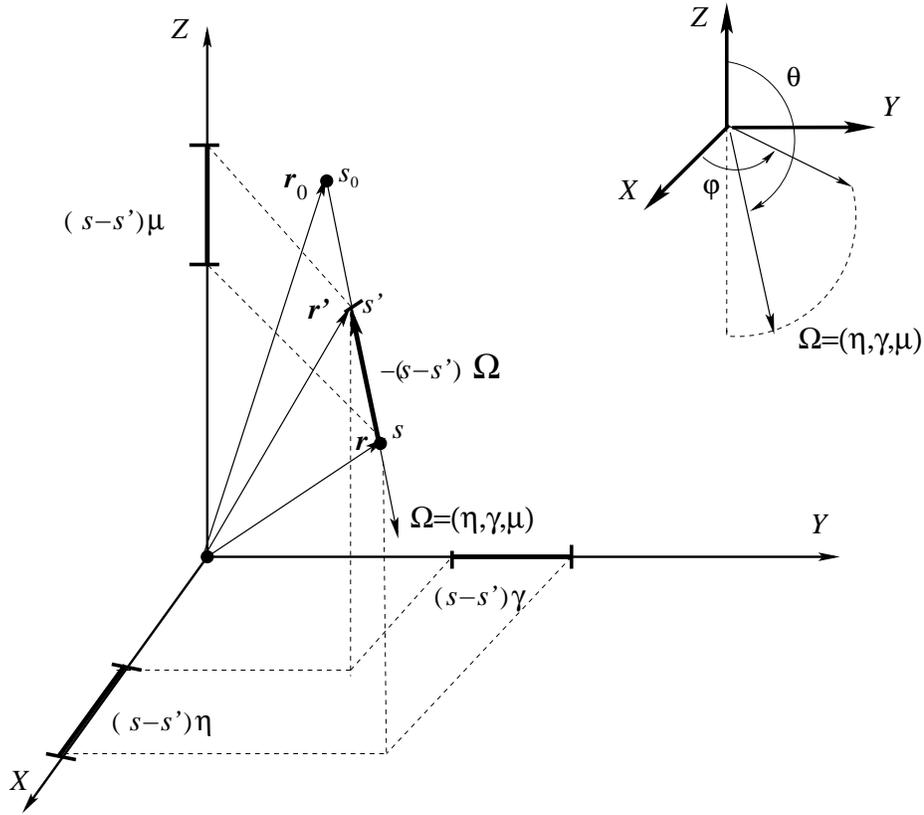}
\caption{The definition of the spatial location $\bm{r}$ and the
projected distances $(s-s')\bm{\Omega}$ which appear in the 2D formal
solution integral (Equation~(\ref{formal-solution})). 
$\bm{r}_0$ and $\bm{r}$ are the initial and final
locations considered in the formal solution integral. 
The values of the variable along the ray satisfy $s_0<s'<s$.
}
\label{fig-fs}
\end{figure*}

The formal solution of Equation~(\ref{rte-reduced}) is given by
\begin{eqnarray}
\!\!\!\!\!\!&&\bm{\mathcal I}(\bm{r}, \bm{\Omega}, x)
=\bm{\mathcal I}(\bm{r}_0, \bm{\Omega}, x)
\ue^{-\displaystyle{\int_{s_0}^{s}} \kappa_{\rm tot}(\bm{r}-(s-s^{\prime})
\bm{\Omega}, x) {\rm d}s^{\prime}}\nonumber \\
\!\!\!\!\!\!&& +\int_{s_0}^{s} 
\bm{\mathcal S}(\bm{r}-(s-s^{\prime})\bm{\Omega}, \bm{\Omega}, x)
\ue^{-\displaystyle{\int_{s'}^{s}}  
\kappa_{\rm tot}(\bm{r}-(s-s^{\prime\prime})
\bm{\Omega},x) {\rm d}s^{\prime\prime}} \nonumber \\
\!\!\!\!\! &&\times [\kappa_{\rm tot}(\bm{r}-(s-s^{\prime})\bm{\Omega},x)]
{\rm d}s^{\prime}.
\label{formal-solution}
\end{eqnarray}
The formal solution can also be expressed as
\begin{eqnarray}
&&\!\!\!\!\!\!\!\!\!\bm{\mathcal{I}}(\bm{r},\bm{\Omega}, x)=
\bm{\mathcal {I}}(\bm{r}_0, \bm{\Omega}, x)
e^{-\tau_{x}(\bm{r},\bm{\Omega})}\nonumber \\
&&\!\!\!\!\!\!\!\!\!
+\int_{0}^{\tau_{x}(\bm{r},\bm{\Omega})} 
e^{-\tau'_{x}(\bm{r}',\bm{\Omega})}
\bm{\mathcal {S}}(\bm{r}',  \bm{\Omega}, x)d\tau'_{x}(\bm{r}',\bm{\Omega}).
\label{i-out-tau}
\end{eqnarray}
Here $\bm{\mathcal {I}}(\bm{r}_0, \bm{\Omega}, x)$ 
is the boundary condition imposed at the boundary point 
$\bm{r}_0=({\rm x}_0,{\rm y}_0,{\rm z}_0)$.
The monochromatic optical depth scale is defined as
\begin{equation}
\tau_x(\bm{r},\bm{\Omega})=
\tau_x({\rm x},{\rm y}, {\rm z},\bm{\Omega})
=\int_{s_0}^s \kappa_{\rm tot}(\bm{r}-(s-s')
\bm{\Omega}, x)\, ds',
\label{tau}
\end{equation}
$\tau_{x}({\bm{r}},\bm{\Omega})$ is the optical thickness 
from the point $\bm{r}_0$ to the point $\bm{r}$ measured along the ray. 
In Figure \ref{fig-fs} we show the construction of the
vector $\bm {r}'=\bm {r} - (s-s')\bm{\Omega}$. The point $\bm {r}'$, tip of
the vector $\bm {r}'$, runs along the ray from the point $\bm {r}_0$ to
the point $\bm {r}$ as the variable along the ray varies from $s_0$ to
$s$. In the preceding papers (I to IV), the figure corresponding to
Figure \ref{fig-fs} was drawn for a ray passing through the origin of the 
coordinate system.

In paper IV we have shown that using Fourier series expansions of the 
AD PRD functions 
$r_{\rm II, III}(x,x',\bm{\Omega},\bm{\Omega'})$ with respect to the
azimuth ($\varphi$) of the scattered ray, we can transform 
Equations~(\ref{rte-reduced})--(\ref{i-out-tau}) into a simplified
set of equations. In the non-magnetic case, the method described in 
Paper IV can be implemented numerically, without any modifications. 
In the magnetic case, it becomes necessary to slightly 
modify that method to avoid making certain approximations
which otherwise would have to be used (see Section~\ref{rteff} for details). 

\subsection{A FOURIER DECOMPOSITION TECHNIQUE FOR DOMAIN BASED PRD}
\label{rteff}
In the presence of a weak magnetic field $\bm{B}$ defined by its strength $B$
and the orientation $(\theta_B,\chi_B)$, the scattering polarization is 
modified through the Hanle effect. 
A general PRD theory including the Hanle effect was developed in 
\citet{bom97a,bom97b}. A description of the Hanle effect
with the AD PRD functions is given by the approximation level II
described in \citet{bom97b}. In this approximation the frequency space
$(x,x')$ is divided into five domains and the functional forms of the 
redistribution matrices is different in each of these domains.
We start with the AD redistribution matrix including Hanle effect
namely
\begin{eqnarray}
\!\!\!\!\!\!&&{\hat{R}}(x, x', \bm{\Omega}, \bm{\Omega}', \bm{B})=\nonumber \\
\!\!\!\!\!\!&&
\hat{W}\Big\{\hat{M}_{\rm II}(\bm{B},x,x')
r_{\rm II}(x, x', \bm{\Omega}, \bm{\Omega'}) \nonumber \\
\!\!\!\!\!\!&&+\hat{M}_{\rm III}(\bm{B},x,x')
r_{\rm III}(x, x',\bm{\Omega}, \bm{\Omega'}) \Big\}
\hat{\Psi}(\bm{\Omega}').
\label{rmat}
\end{eqnarray}
We recall here that the dependence of the matrices $\hat{M}_{\rm II,III}$
on $x$ and $x'$ is related to the definition of the frequency domains.
Here $\hat{R}$ is a $6\times6$ matrix.
The Fourier series expansions of the functions 
$r_{\rm II, III}(x, x', \bm{\Omega}, \bm{\Omega'})$
is written as
\begin{eqnarray}
&&r_{\rm II,III}(x,x',\bm{\Omega},\bm{\Omega'})=
\nonumber\\
&&\sum_{k=0}^{k=\infty}\,\, (2-\delta_{k0})
e^{ik\varphi}\,\,{\tilde{r}}_{\rm II,III}^{(k)}(x,x',\theta,\bm{\Omega}'),
\label{Fourier-series-r23-pf}
\end{eqnarray}
with
\begin{eqnarray}
&&{\tilde{r}_{\rm II,III}}^{(k)}(x,x',\theta,\bm{\Omega'})
=\int_0^{2\pi}\frac{d\,\varphi}{2\pi}\, e^{-ik\varphi}\,\,
\nonumber\\
&&\times {r}_{\rm II,III}(x, x', \bm{\Omega}, \bm{\Omega'}).
\nonumber \\
\label{f-r-coefficients}
\end{eqnarray}

Applying this expansion we can derive a polarized RT equation 
in terms of the Fourier coefficients $\tilde{\bm{\mathcal I}}^{(k)}$ and 
$\tilde{\bm{\mathcal S}}^{(k)}$ (see Section~4 of 
Paper IV for details) namely
\begin{eqnarray}
&&-\frac{1}{\kappa_{\rm tot}(\bm{r}, x)}\bm{\Omega} \cdot
\bm{\nabla}{\tilde{\bm{\mathcal{I}}}}^{(k)}
(\bm{r}, \bm{\Omega}, x) = \nonumber \\
\!\!\!\!\!\!&&[{\tilde{\bm{\mathcal{I}}}}^{(k)}
(\bm{r}, \bm{\Omega}, x)-
{\tilde{\bm{\mathcal S}}}^{(k)}(\bm{r}, \theta, x)],
\label{f-rte}
\end{eqnarray}
where
\begin{eqnarray}
&&\bm{\mathcal S}(\bm{r}, \bm{\Omega}, x)=\nonumber\\
&&\sum_{k=0}^{k=\infty}(2-\delta_{k0})
\Big\{\cos(k\varphi){\mathcal Re}\left[
{\tilde{\bm{\mathcal {S}}}^{(k)}}(\bm{r}, \theta, x)\right]\nonumber\\
&&-\sin(k\varphi){\mathcal Im}\left[{\tilde{\bm{\mathcal {S}}}^{(k)}}
(\bm{r}, \theta, x)\right] \Big\},\nonumber\\
\label{transform-1f}
\end{eqnarray}
and
\begin{eqnarray}
&&\bm{\mathcal I}(\bm{r}, \bm{\Omega}, x)=\nonumber\\
&&\sum_{k=0}^{k=\infty}(2-\delta_{k0})\Big\{\cos(k\varphi){\mathcal Re}
\left[{\tilde{\bm{\mathcal {I}}}^{(k)}}(\bm{r}, \bm{\Omega}, x)\right]
\nonumber\\
&&-\sin(k\varphi){\mathcal Im}\left[{\tilde{\bm{\mathcal {I}}}^{(k)}}
(\bm{r}, \bm{\Omega}, x)\right] \Big\}.\nonumber\\
\label{transform-2f}
\end{eqnarray}

Equation~(\ref{f-rte}) represents the most reduced form of polarized
RT equation in multi-D geometry with the AD PRD.
Hereafter we refer to $\tilde{\bm{\mathcal I}}^{(k)}$ and
$\tilde{\bm{\mathcal S}}^{(k)}$ as `irreducible Fourier coefficients'.
$\tilde{\bm{\mathcal I}}^{(k)}$ and 
$\tilde{\bm{\mathcal S}}^{(k)}$ are 6-dimensional complex vectors for 
each value of $k$.
Here 
\begin{eqnarray}
&&{\tilde{\bm{\mathcal {S}}}^{(k)}}(\bm{r}, \theta, x)=
p_x{\tilde{\bm{\mathcal {S}}}_l^{(k)}}(\bm{r}, \theta, x)
\nonumber\\
&&+(1-p_x){\tilde{\bm{\mathcal S}}}_C^{(k)}(\bm{r},x),
\label{f-stot-new}
\end{eqnarray}
with
\begin{equation}
{\tilde{\bm{\mathcal S}}}_C^{(k)}(\bm{r},x)=\delta_{k0} 
\bm{\mathcal S}_C(\bm{r},x),
\label{f-sc-new}
\end{equation}
and
\begin{eqnarray}
\!\!\!\!\!\!&&{\tilde{\bm{\mathcal {S}}}_l^{(k)}}(\bm{r}, \theta, x)=
{\tilde{\bm{\mathcal G}}}^{(k)}(\bm{r})
+\frac{1}{\phi(x)} \int_{-\infty}^{+\infty} dx' \nonumber \\
\!\!\!\!\!\!&&
\times\oint\frac{d \bm{\Omega}'} {4 \pi} 
{\hat{\tilde{R}}}^{(k)}(x, x', \theta, \bm{\Omega}', \bm{B})\nonumber \\
\!\!\!\!\!\!&&\times \sum_{k'=0}^{k'=+\infty}e^{ik'\varphi'}  (2-\delta_{k'0})
{\tilde{\bm{\mathcal{I}}}}^{(k')}(\bm{r}, \bm{\Omega}',x').
\label{f-sl-i-new}
\end{eqnarray}
Here ${\tilde{\bm{\mathcal G}}}^{(k)}(\bm{r})=\epsilon 
\delta_{k0} B_{\nu}(\bm{r})$ and
\begin{eqnarray}
\!\!\!\!\!\!&&{\hat{\tilde{R}}}^{(k)}(x, x', \theta, \bm{\Omega}', \bm{B})=
\nonumber \\
\!\!\!\!\!\!&&
\hat{W}\Big\{\hat{M}_{\rm II}(\bm{B},x,x')
{\tilde{r}}^{(k)}_{\rm II}(x, x', \theta, \bm{\Omega'}) \nonumber \\
\!\!\!\!\!\!&&+\hat{M}_{\rm III}(\bm{B},x,x')
{\tilde{r}}^{(k)}_{\rm III}(x, x',\theta, \bm{\Omega'}) \Big\}
\hat{\Psi}(\bm{\Omega}').
\label{rmat-f}
\end{eqnarray}
Clearly, in the above equation the matrix 
${\hat{\tilde{R}}}^{(k)}$ is independent
of the azimuth $(\varphi)$ of the scattered ray. 
We recall that $\hat{M}_{\rm II, III}$ matrices have different
forms in different frequency domains 
\citep[see][and Appendix~A of Anusha et al. 2011b]{bom97b,knnetal02}.
In the approximation level--II of \citet{bom97b} 
the expressions for the frequency
domains depend on the scattering angle $\Theta$, and hence on
$\bm{\Omega}$ and $\bm{\Omega}'$ (because $\cos \Theta = 
\bm{\Omega} \cdot \bm{\Omega}'$). Therefore to be consistent, we 
must apply the Fourier series expansions to the functions involving 
$\Theta$ which appear in the statements defining the 
AD frequency domains of \citet{bom97b}.
This leads to complicated mathematical forms of the domain statements.
To a first approximation one can keep only the dominant term in the Fourier
series (corresponding to the term with $k=0$). This amounts to replacing
the AD frequency domain expressions by their 
azimuth ($\varphi$)-averages. A similar averaging of the
domains over the variable ($\varphi-\varphi'$) is done 
in \citet{msknn11}, where the authors solve the Hanle RT problem with
the AD PRD in 1D planar geometry.
These kinds of averaging can lead to loss of some information on the
azimuth ($\varphi$) dependence of the scattered ray in the domain expressions. 
A better and alternative approach which avoids any averaging of the domains
is the following.

Substituting Equation~(\ref{Fourier-series-r23-pf}) in Equation~(\ref{rmat})
we can write the $ij$-th element of the ${\hat{R}}$ matrix as
\begin{eqnarray}
\!\!\!\!\!\!&&{R}_{ij}(x, x', \bm{\Omega}, \bm{\Omega}', \bm{B})=
\nonumber \\
\!\!\!\!\!\!&&
\sum_{k=0}^{k=\infty}\,\, (2-\delta_{k0})
e^{ik\varphi}\,\,
{\tilde{R}}_{ij}^{(k)}(x, x', \theta, \bm{\Omega}', \bm{B}),\nonumber \\
&&\quad i,j=1,2,\ldots,6,
\label{f-exp-rmat-ij}
\end{eqnarray}
with $\tilde{R}_{ij}^{(k)}$ being the elements of the matrix 
${\hat{\tilde{R}}}^{(k)}$ given by Equation~(\ref{rmat-f}). 
Through the $2\pi$-periodicity of the redistribution functions
$r_{\rm II, III}(x, x', \bm{\Omega}, \bm{\Omega'})$ each element
of the ${\hat{R}}$ matrix becomes $2\pi$-periodic.
Therefore we can identify that Equation~(\ref{f-exp-rmat-ij}) represents the 
Fourier series expansion of the elements ${R}_{ij}$ of the ${\hat{R}}$ matrix,
with ${{\tilde{R}}}_{ij}^{(k)}$ being the Fourier coefficients.
Thus, instead of computing ${\hat{\tilde{R}}}^{(k)}$ using 
Equation~(\ref{rmat-f}) it is advantageous to compute its elements 
through the definition of the Fourier coefficients, namely
\begin{eqnarray}
&&{\tilde{R}}_{ij}^{(k)}(x,x',\theta,\bm{\Omega'})
=\int_0^{2\pi}\frac{d\,\varphi}{2\pi}\, e^{-ik\varphi}\,\,
{W}_{ij}
\nonumber\\
&&\Big\{({M}_{\rm II})_{ij}(\bm{B},x,x')
{r}_{\rm II}(x, x', \bm{\Omega}, \bm{\Omega'}) \nonumber \\
\!\!\!\!\!\!&&+({M}_{\rm III})_{ij}(\bm{B},x,x')
{r}_{\rm III}(x, x',\bm{\Omega}, \bm{\Omega'}) \Big\}.
\label{f-coefficients}
\end{eqnarray}
Here $W_{ij}$ are the elements of the $\hat{W}$ matrix and 
the matrix elements $(\hat{M}_{\rm II, III})_{ij}$ are computed using
the AD expressions for the frequency domains
as done in \citet{knnetal02}, without performing azimuth 
averaging of the domains.

\section{NUMERICAL METHOD OF SOLUTION}
\label{numerics}
A fast iterative method called the preconditioned Stabilized 
Bi-Conjugate Gradient (Pre-BiCG-STAB) was developed 
for 2D transfer with PRD in Paper II. Non-magnetic 2D slabs and the AA PRD
were considered in that paper.
An extension to a magnetized 3D medium with the AA 
PRD was taken up in Paper-III. 
In all these papers, the computing algorithm was written in the 
$n$-dimensional Euclidean space of real numbers ${\mathbb{R}}^n$. 
In the present paper, we extend the method to handle the AD PRD 
for a magnetized 2D media. In this case, it is advantageous to formulate 
the computing algorithm
in the $n$-dimensional complex space ${\mathbb{C}}^n$. 
Here $n=n_k \times n_p \times n_{\theta} \times n_x \times n_Y \times n_Z$, 
where $n_{Y,Z}$ are
the number of grid points in the $Y$ and $Z$ directions, and $n_x$ refers
to the number of frequency points. $n_{\theta}$ is the number of polar 
angles ($\theta$) considered in the problem. $n_p$ is the number of polarization
components of the irreducible vectors. $n_p=6$ for both non-magnetic 
and magnetic AD PRD cases. $n_k$ is the number of components 
retained in the Fourier series expansions of the AD 
PRD functions. 
Based on the studies in Paper IV we take $n_k=5$. Clearly the dimensionality
of the problem increases when we handle the AD PRD in line scattering
in comparison with the AA PRD (see Papers II and III).
The numerical results presented in this paper correspond to 2D media.
For 3D RT, the dimensionality escalates, and it is more computationally 
demanding than the 2D RT. The computing algorithm is similar to the one
given in Paper II, with straightforward extensions to handle the AD
PRD. The essential difference is that we now use the vectors in the complex
space $\mathbb{C}^n$. The algorithm contains operations involving
the inner product $\la\,,\ra$.
In ${\mathbb{C}}^n$ the inner product of two vectors 
$\bm{u}=(u_1,u_2,\cdot\cdot\cdot,u_n)^T$ and 
$\bm{v}=(v_1,v_2,\cdot\cdot\cdot,v_n)^T$
is defined as
\begin{equation}
\la \bm{u}, \bm{v} \ra = \sum_{i=1}^n u_i v_i^*,
\label{inner}
\end{equation}
where $*$ represents complex conjugation.\\

\noindent
{\it The Preconditioner matrix}\\

\noindent
The preconditioner matrices are any form of implicit or explicit
modification of the original matrix in the system of 
equations to be solved, which accelerate the rate of convergence
of the problem \citep[see][]{saad00}. As explained in Paper III,
the magnetic case requires the use of domain based PRD, where
it becomes necessary to use different preconditioner matrices in different
frequency domains.  In the problem under consideration the preconditioner 
matrices are complex block diagonal matrices. The dimension
of each block is $n_x \times n_x$, and the total number of such blocks 
is $n/n_x$. The construction of the preconditioner matrices is analogous
to that described in Paper III, with the appropriate modifications
to handle the Fourier decomposed AD PRD matrices. 

\section{RESULTS AND DISCUSSIONS}
\label{results}
In this section we study some of the benchmark results obtained using the 
method proposed in this paper (Sections~\ref{rteff} and \ref{numerics})
which is based on the Fourier decomposition technique developed
in Paper IV. 
In all the results, we consider the following
global model parameters.
The damping parameter
of the Voigt profile is $a=2 \times 10^{-3}$
and the continuum to the line opacity $\kappa_c/\kappa_l=10^{-7}$.
The internal thermal sources are taken as constant (the Planck
function $B_{\nu}(\bm{r})=1$). The medium
is assumed to be isothermal and self-emitting (no incident radiation on the
boundaries). 
The ratios of elastic and inelastic collision rates to the radiative
de-excitation rate are respectively
$\Gamma_E/\Gamma_R=10^{-4}$, $\Gamma_I/\Gamma_R=10^{-4}$.
The expressions for the redistribution matrices
contain the parameters $\alpha$ and $\beta^{(K)}$ and
are called as branching ratios \citep[see][]{bom97b}.
They are defined as
\begin{equation}
\alpha=\frac{\Gamma_R}{\Gamma_R+\Gamma_E+\Gamma_I},
\label{alp}
\end{equation}

\begin{equation}
\beta^{(K)}=\frac{\Gamma_R}{\Gamma_R+D^{(K)}+\Gamma_I},
\label{beta0}
\end{equation}
with $D^{(0)}=0$ and $D^{(2)}=c \Gamma_E$, where $c$ is a constant,
taken to be 0.379 \citep[see][]{mf92}.
The branching ratios for the chosen values of 
$\Gamma_E/\Gamma_R$, $\Gamma_I/\Gamma_R$ and $D^{(K)}$
are $(\alpha,\beta^{(0)},\beta^{(2)}) = (1, 1, 1)$.
They correspond to a PRD scattering matrix that 
uses only ${\tilde{r}}^{(k)}_{\rm II}(x,x',\theta,\bm{\Omega'})$
function. In other words we consider only the collisionless redistribution
processes. We parameterize the magnetic field by 
$(\Gamma_B,\theta_B,\chi_B)$. 
The Hanle $\Gamma_B$ coefficient \citep[see][]{bom97b} takes
two different forms, namely
\begin{equation}
\Gamma_B=\Gamma'_K=\beta^{(K)}\Gamma,\quad\Gamma_B=\Gamma''=\alpha\Gamma,
\label{gamma}
\end{equation}
with
\begin{equation}
\Gamma=g_J\, \frac{2\pi eB}{2m_e\Gamma_R}
\end{equation}
where ${eB}/{2m_e}$ is the Larmor frequency
of the electron in the magnetic field (with $e$ and $m_e$ being
the charge and mass of the electron). We take
$\Gamma_B=1$ for computing all the results presented
in Section~\ref{results}. 
In this paper we restrict our attention to effectively
optically thin cases (namely the optical thicknesses $T_Y=T_Z=20$). 
They represent formation of weak resonance lines in finite
dimensional structures. Studies on the effects of the AD 
PRD in optically thick lines is deferred
to a later paper. 

We show the relative importance of the 
AD PRD in comparison with
the AA PRD considering 
(1) non-magnetic case ($\bm{B}=0$), and (2) magnetic case ($\bm{B}\ne0$).

\begin{figure*}
\centering
\includegraphics[scale=0.4]{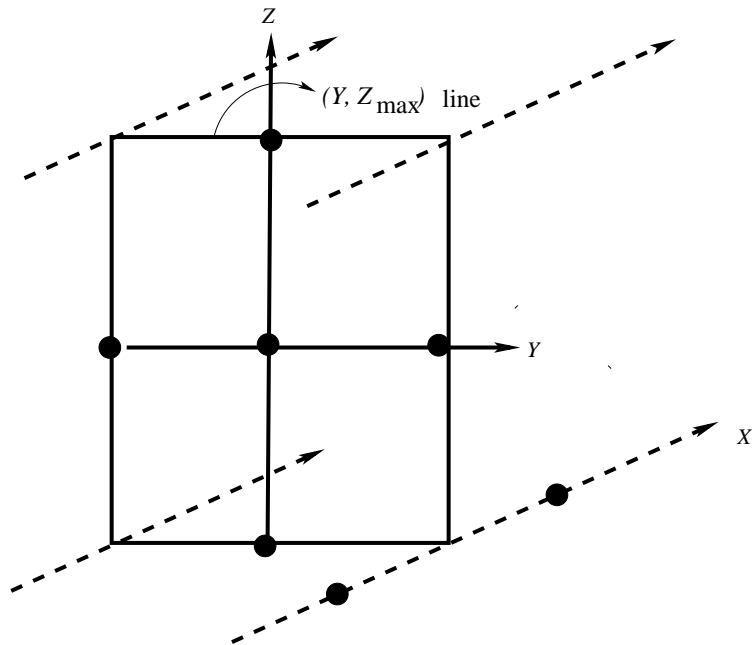}
\caption{RT in a 2D medium. We assume that the medium is infinite in the
direction of the $X$-axis and has a finite dimension in the direction of 
the $Y$-axis and the $Z$-axis. The top surface is marked.
}
\label{fig-2d-geometry}
\end{figure*}

In Figure~\ref{fig-2d-geometry} we show the geometry 
of RT in a 2D medium.
We assume that the medium is infinite along the $X$-axis, 
and finite along the $Y$- and $Z$-axes. 
The top surface of the 2D medium is defined to be the line
$(Y,Z_{\rm max})$, as marked in Figure~\ref{fig-2d-geometry}. We obtain the 
emergent, spatially averaged $(I,Q/I,U/I)$ profiles, by simply performing the 
arithmetic average of these profiles over this line $(Y,Z_{\rm max})$
on the top surface.

\subsection{Nature of the components of $\bm{\mathcal I}$ and 
$\tilde{\bm{\mathcal I}}^{(k)}$}
Often it is pointed out in the literature that the AD PRD effects
are important \citep[see e.g.,][]{knnetal02} for polarized
line formation. For multi-D polarized RT the AD PRD effects have not been
addresses so far. Therefore we would like to quantitatively examine this
aspect by taking the example of polarized line formation in 2D media, through 
explicit computation of Stokes profiles using the AD and the AA 
PRD mechanisms for both $\bm{B}=0$ and $\bm{B}\ne0$ cases.
The Stokes parameters $Q$ and $U$ contain inherently all the
AD PRD informations. In order to understand the actual differences
between the AD and the AA solutions one has to study the frequency and
angular behaviour of the more fundamental quantities, namely
$\bm{\mathcal I}$ and $\tilde{\bm{\mathcal I}}^{(k)}$, which are
obtained through multi-polar expansions of the Stokes
parameters.

\begin{figure*}
\centering
\includegraphics[scale=0.7]{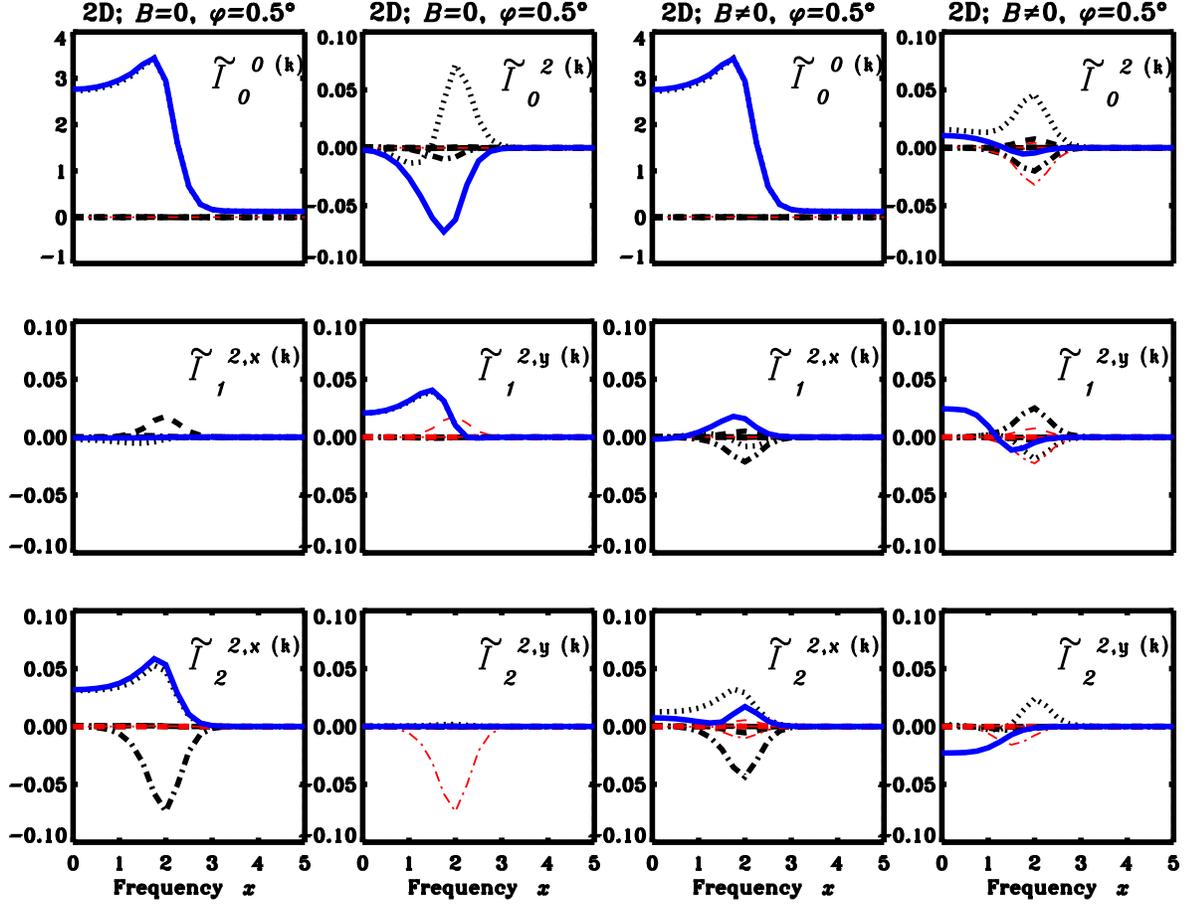}
\caption{The emergent, spatially averaged components of 
$\tilde{\bm{\mathcal I}}^{(k)}$ in non-magnetic (the first two 
columns) and magnetic (the last two columns) 2D media for $\mu=0.11$ and 
$\varphi=89^{\circ}$. The actual values of the components are scaled up
by a factor of $10^4$. Solid lines represent the 
components of $\bm{\mathcal I}$ for the AA PRD, plotted here for comparison. 
The dotted curves represent the components $\tilde{\bm{\mathcal I}}^{(0)}$. 
The thick curves with dashed, dot-dashed, dash-triple-dotted and long-dashed 
line types respectively represent 
${\mathcal Re}\left[\tilde{\bm{\mathcal{I}}}^{(1)}\right]$, 
${\mathcal Re}\left[\tilde{\bm{\mathcal{I}}}^{(2)}\right]$, 
${\mathcal Re}\left[\tilde{\bm{\mathcal{I}}}^{(3)}\right]$ and 
${\mathcal Re}\left[\tilde{\bm{\mathcal{I}}}^{(4)}\right]$. Similarly
the thin curves with dashed, dot-dashed, dash-triple-dotted and long-dashed 
line types respectively represent 
${\mathcal Im}\left[\tilde{\bm{\mathcal{I}}}^{(1)}\right]$,    
${\mathcal Im}\left[\tilde{\bm{\mathcal{I}}}^{(2)}\right]$, 
${\mathcal Im}\left[\tilde{\bm{\mathcal{I}}}^{(3)}\right]$ and   
${\mathcal Im}\left[\tilde{\bm{\mathcal{I}}}^{(4)}\right]$.
In the last two columns, the
magnetic field parameters are $(\Gamma_B,\theta_B,\chi_B)
=(1,90^{\circ},60^{\circ})$.}
\label{fig-I-1to6-nonmag-fourier}
\end{figure*}

\begin{figure*}
\centering
\includegraphics[scale=0.7]{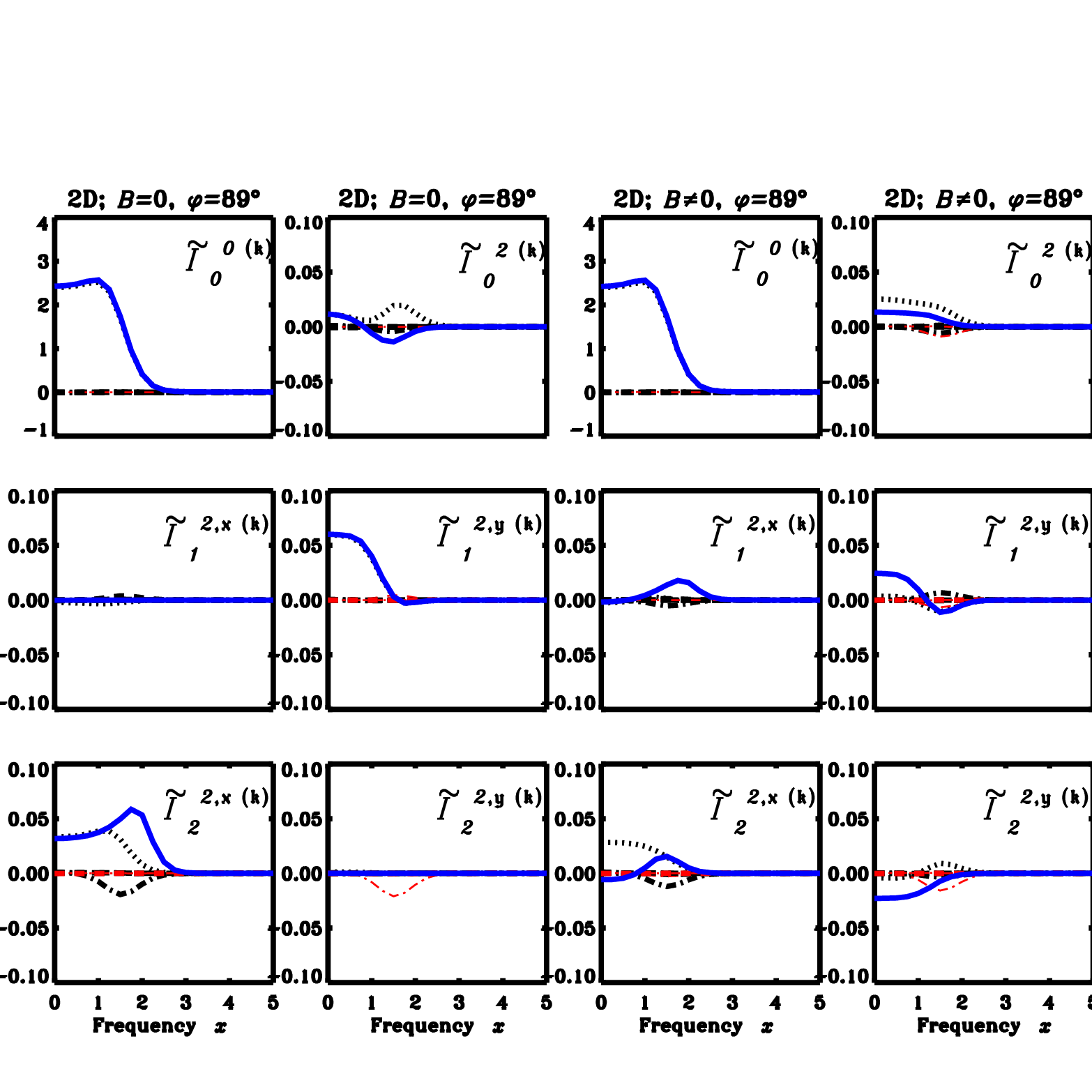}
\caption{Same as Figure~\ref{fig-I-1to6-nonmag-fourier} but for 
$\varphi=89^{\circ}$.}
\label{fig-I-1to6-mag-fourier}
\end{figure*}

In Figures~\ref{fig-I-1to6-nonmag} and \ref{fig-I-1to6-mag}, we plot the 
components of the real vector $\bm{\mathcal I}$=($I^0_0$, $I^2_0$, 
$I^{2,\rm x}_1$, $I^{2,\rm y}_1$, $I^{2,\rm x}_2$, $I^{2,\rm y}_2$) which are
constructed using the 6 irreducible components of the nine vectors 
$\tilde{\bm{\mathcal I}}^{(0)}$, 
${\mathcal Re}\left[\tilde{\bm{\mathcal I}}^{(1)}\right]$,
${\mathcal Im}\left[\tilde{\bm{\mathcal I}}^{(1)}\right]$,
${\mathcal Re}\left[\tilde{\bm{\mathcal I}}^{(2)}\right]$,
${\mathcal Im}\left[\tilde{\bm{\mathcal I}}^{(2)}\right]$,
${\mathcal Re}\left[\tilde{\bm{\mathcal I}}^{(3)}\right]$,
${\mathcal Im}\left[\tilde{\bm{\mathcal I}}^{(3)}\right]$,
${\mathcal Re}\left[\tilde{\bm{\mathcal I}}^{(4)}\right]$ and
${\mathcal Im}\left[\tilde{\bm{\mathcal I}}^{(4)}\right]$.
For each $k$, $\tilde{\bm{\mathcal I}}^{(k)}$ is a 6-component
complex vector ($\tilde{I}^{0\,(k)}_0$, $\tilde{I}^{2\,(k)}_0$, 
$\tilde{I}^{2,\rm x\,(k)}_1$, $\tilde{I}^{2,\rm y\,(k)}_1$, 
$\tilde{I}^{2,\rm x\,(k)}_2$, $\tilde{I}^{2,\rm y\,(k)}_2$).
Thus in Figures~\ref{fig-I-1to6-nonmag-fourier}
and \ref{fig-I-1to6-mag-fourier} there are 54 components 
plotted in 6 panels, with each panel containing 9 curves (see the caption of 
Figure~\ref{fig-I-1to6-nonmag-fourier} for line identifications).
In the Figures~\ref{fig-I-1to6-nonmag}, \ref{fig-I-1to6-mag},
\ref{fig-I-1to6-nonmag-fourier} and \ref{fig-I-1to6-mag-fourier} 
the first two columns correspond to the $\bm{B}=0$ case
and the last two columns correspond to the $\bm{B}\ne0$ case.
Here we have chosen $\mu=0.11$ and 
two examples of $\varphi$ namely $0.5^{\circ}$
and $89^{\circ}$. 
$\bm{\mathcal I}$ and $\tilde{\bm{\mathcal I}}^{(k)}$ are related
through Equation~(\ref{transform-2f}) which can be re-written 
by truncating the Fourier series to five terms, as discussed and
validated in Paper IV.
Equation~(\ref{transform-2f}) can be approximated by
\begin{eqnarray}
\bm{\mathcal I}\approx{\tilde{\bm{\mathcal {I}}}^{(0)}}
+\sum_{k=1}^{k=4}2\,{\mathcal Re}\left[
{\tilde{\bm{\mathcal {I}}}^{(k)}}\right],
\label{transform-2f1}
\end{eqnarray}
for $\varphi=0.5^{\circ}$ and
\begin{eqnarray}
&&\bm{\mathcal I}\approx{\tilde{\bm{\mathcal {I}}}^{(0)}}
-2\,\Bigg\{ 
{\mathcal Im}\left[\tilde{\bm{\mathcal {I}}}^{(1)}\right] 
+{\mathcal Re}\left[\tilde{\bm{\mathcal {I}}}^{(2)}\right] \nonumber \\
&&-{\mathcal Im}\left[\tilde{\bm{\mathcal {I}}}^{(3)}\right] 
-{\mathcal Re}\left[\tilde{\bm{\mathcal {I}}}^{(4)}\right] 
\Bigg \},
\label{transform-2f2}
\end{eqnarray}
for $\varphi=89^{\circ}$.
\subsubsection{Non-magnetic case}
\begin{figure*}
\centering
\includegraphics[scale=0.7]{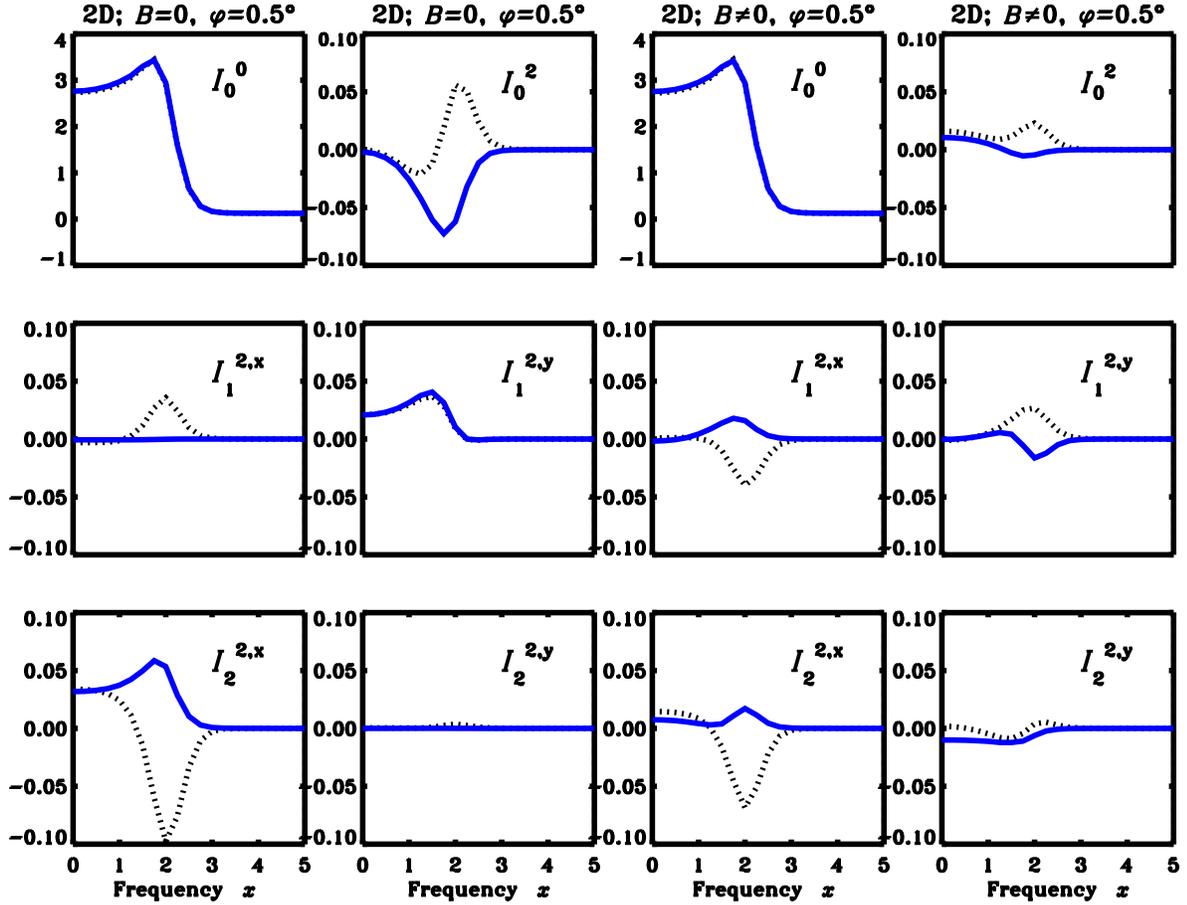}
\caption{The emergent, surface averaged components of $\bm{\mathcal I}$ in 
non-magnetic (the first two columns) and magnetic (the last two 
columns) 2D media for $\mu=0.11$ and $\varphi=0.5^{\circ}$.
The actual values of the components are scaled up by a factor of $10^4$.
Solid and dotted lines represent respectively the AA and the AD PRD. 
In the first two columns (for $\bm{B}=0$), 
$I^{2,\rm x}_1$ and $I^{2,\rm y}_2$ are zero for the AA PRD (solid 
lines) and the other 10 components are non-zero (four AA components 
and six AD components). In the last two columns, the 
magnetic field parameters are $(\Gamma_B,\theta_B,\chi_B)
=(1,90^{\circ},60^{\circ})$. All the components are important for 
$\bm{B}\ne0$.}
\label{fig-I-1to6-nonmag}
\end{figure*}

In general the component $I^0_0$ (and hence Stokes $I$ parameter) 
is less sensitive to the AD nature of PRD functions. Only for certain
choices of $(\theta,\varphi)$, does $[I^0_0]_{\rm AD}$ differ noticeably
from $[I^0_0]_{\rm AA}$. The other polarization components exhibit
significant sensitivity to the AD PRD. For the present choice of
$(\theta,\varphi)$, in the second column of 
Figure~\ref{fig-I-1to6-nonmag} we see 
that $[I^{2,\rm y}_1]_{\rm AD}$ and
$[I^{2,\rm y}_1]_{\rm AA}$ are nearly the same. We have verified that
they differ very much for other choices of $(\theta,\varphi)$.
Thus the differences between the AD PRD and the AA PRD are disclosed only when 
we consider polarization components and not just the $I^0_0$ component.

In the following we discuss the important symmetry relations of the
polarized radiation field for a non-magnetic 2D medium.

\begin{figure*}
\centering
\includegraphics[scale=0.7]{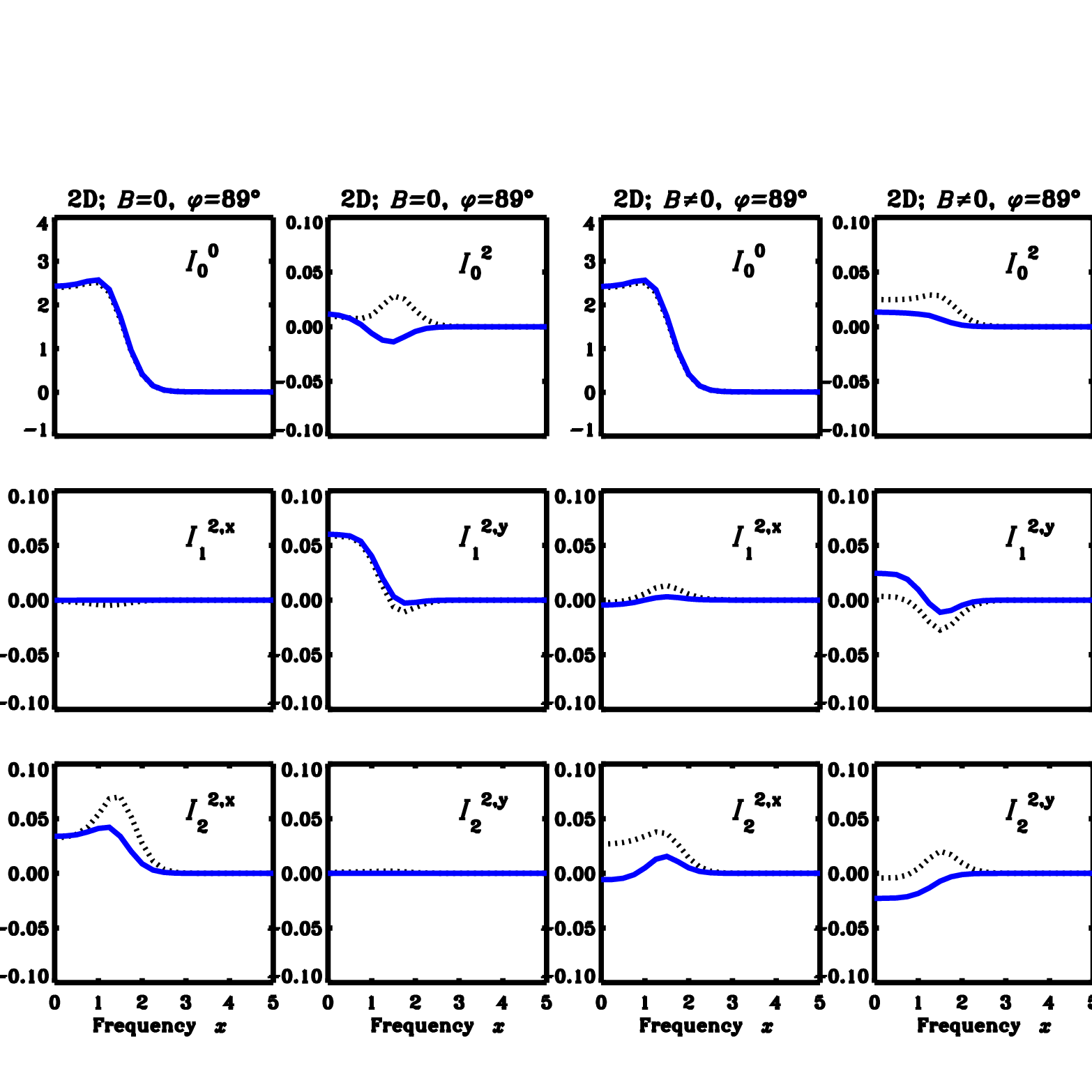}
\caption{Same as Figure~\ref{fig-I-1to6-nonmag} but for $\varphi=89^{\circ}$.}
\label{fig-I-1to6-mag}
\end{figure*}

\subsubsubsection{Symmetry relations in non-magnetic 2D media}
In Paper II we have shown that $[I^{2, \rm x}_1]_{\rm AA}$ and
$[I^{2, \rm y}_2]_{\rm AA}$ are identically zero in non-magnetic 2D media 
(shown as solid lines in the first two columns of
Figures~\ref{fig-I-1to6-nonmag} and 
\ref{fig-I-1to6-mag}). This property of $I^{2,\rm x}_1$ and
$I^{2,\rm y}_2$ in a non-magnetic 2D medium arises from the symmetry of
the Stokes $I$ parameter with respect to the infinite axis of the medium 
($X$-axis in our case), combined with the $\varphi$-dependence of the 
geometrical factors ${\mathcal T}^K_Q(i,\bm{\Omega})$ (see Appendix B of 
Paper II, Equations~(B9) and (B10)). Such a symmetry property is valid 
if the scattering is according to CRD or the AA PRD where 
the angular dependence of the source vectors occurs only through the 
angular dependence of $(I,Q,U)$ and that of ${\mathcal T}^K_Q(i,\bm{\Omega})$. 
For the AD PRD, in addition to these two factors, 
the angle-dependence of the PRD functions also causes change in the 
angular behaviour of the source vectors. Thus the AD 
$r_{\rm II,III}$ functions depend on $\varphi$ in such a way
that $[I^{2,\rm x}_1]_{\rm AD}$ and $[I^{2,\rm y}_2]_{\rm AD}$ 
are not zero in general (shown as dotted lines in the first two columns of 
Figures~\ref{fig-I-1to6-nonmag} and \ref{fig-I-1to6-mag}). 
Using a Fourier expansion of the AD $r_{\rm II,III}$ 
functions we have proved this fact in Appendix~\ref{appendixa}.

The components of $\tilde{\bm{\mathcal {I}}}^{(k)}$ also exhibit some 
interesting properties. In Table~\ref{table_1} we list the dominant
Fourier components contributing to each of the 6 components of 
$\bm{\mathcal I}$ in a non-magnetic 
2D medium (shown as crosses).
In the following we describe the nature of these Fourier components. 
Of all the components $\tilde{I}^{0(k)}_0$ and $\tilde{I}^{2(k)}_0$, 
only $\tilde{I}^{0(0)}_0$ and $\tilde{I}^{2(0)}_0$ (dotted lines in 
the first two columns of Figures~\ref{fig-I-1to6-nonmag-fourier} and
\ref{fig-I-1to6-mag-fourier}) are dominant and they are 
nearly same as $[I^0_0]_{\rm AD}$ and $[I^2_0]_{\rm AD}$ respectively 
(dotted lines in the first two columns of 
Figures~\ref{fig-I-1to6-nonmag} and \ref{fig-I-1to6-mag}).
$\tilde{I}^{2(0)}_0$ is an important ingredient for Stokes $Q$. The components
$\tilde{I}^{2,{\rm x,y}(k)}_{1,2}$ are ingredients for both Stokes $Q$ and $U$.
It can be seen that
except $\tilde{I}^{2,{\rm y}(0)}_{2}$ all other 
$\tilde{I}^{2,{\rm x,y}(0)}_{1,2}$ play an important role in the construction 
of the vector $\bm{\mathcal I}$. For $\tilde{I}^{2,{\rm x}(k)}_{1,2}$, 
$k\ne0$, only ${\mathcal Re}\left[\tilde{I}^{2,{\rm x}(1)}_{1}\right]$ and 
${\mathcal Re}\left[\tilde{I}^{2,{\rm x}(2)}_{2}\right]$ (thick dashed 
and thick dot-dashed lines respectively) are dominant. For 
$\tilde{I}^{2,{\rm y}(k)}_{1,2}$, $k\ne0$, only 
${\mathcal Im}\left[\tilde{I}^{2,{\rm y}(1)}_{1}\right]$ 
and ${\mathcal Im}\left[\tilde{I}^{2,{\rm y}(2)}_{2}\right]$ (thin dashed 
and thin dot-dashed lines respectively) are dominant. This property is true for
other choices of $(\theta,\varphi)$ also. From this property it appears
that, in rapid computations involving the AD PRD mechanisms, it may prove
useful to approximate the problem by using the truncated, 
9-component vector ($\tilde{I}^{0(0)}_0$, $\tilde{I}^{2(0)}_0$,
$\tilde{I}^{2,\rm x(0)}_1$, $\tilde{I}^{2,\rm y(0)}_1$, 
${\mathcal Re}\left[\tilde{I}^{2,{\rm x}(1)}_{1}\right]$, 
${\mathcal Im}\left[\tilde{I}^{2,{\rm y}(1)}_{1}\right]$, 
${\mathcal Re}\left[\tilde{I}^{2,{\rm x}(2)}_{2}\right]$, 
${\mathcal Im}\left[\tilde{I}^{2,{\rm y}(2)}_{2}\right]$) 
and obtain sufficiently accurate solution with less computational efforts. 
When the 6-component complex vector $\tilde{\bm{\mathcal I}}^{(k)}$ 
for each value of $k=0,1,2,3,4$, having 54 independent components is used, 
the computations are expensive. 

\subsubsection{Magnetic case}
When we introduce a non-zero magnetic field $\bm{B}$, the shapes, signs and 
magnitudes of $\bm{\mathcal I}_{\rm AA,AD}$ change (see the last two columns of 
Figures~\ref{fig-I-1to6-nonmag} and \ref{fig-I-1to6-mag}). 
$[I^{2,\rm x}_1]_{\rm AA}$ and $[I^{2,\rm y}_2]_{\rm AA}$ which were zero 
when $\bm{B}=0$, now take non-zero values. With a given $\bm{B}\ne0$, 
except $I^0_0$, 
the behaviors of all the other components for the AD PRD are 
very different from those for the AA PRD. Because the Hanle effect 
is operative only in the line core $(0\le x \le 3.5)$, all the magnetic 
effects are confined to the line core. 

For $\bm{B}=0$ only some of the components of 
$\tilde{\bm{\mathcal I}}^{(k)}$ play a significant role. For $\bm{B}\ne0$, all 
the components of $\tilde{\bm{\mathcal I}}^{(k)}$ can become important (see 
the last two columns of Figures~\ref{fig-I-1to6-nonmag-fourier} and 
\ref{fig-I-1to6-mag-fourier}).
This property has a direct impact on the values of $Q/I$ and $U/I$.

\subsection{Emergent Stokes Profiles}
\begin{figure*}
\centering
\includegraphics[scale=0.7]{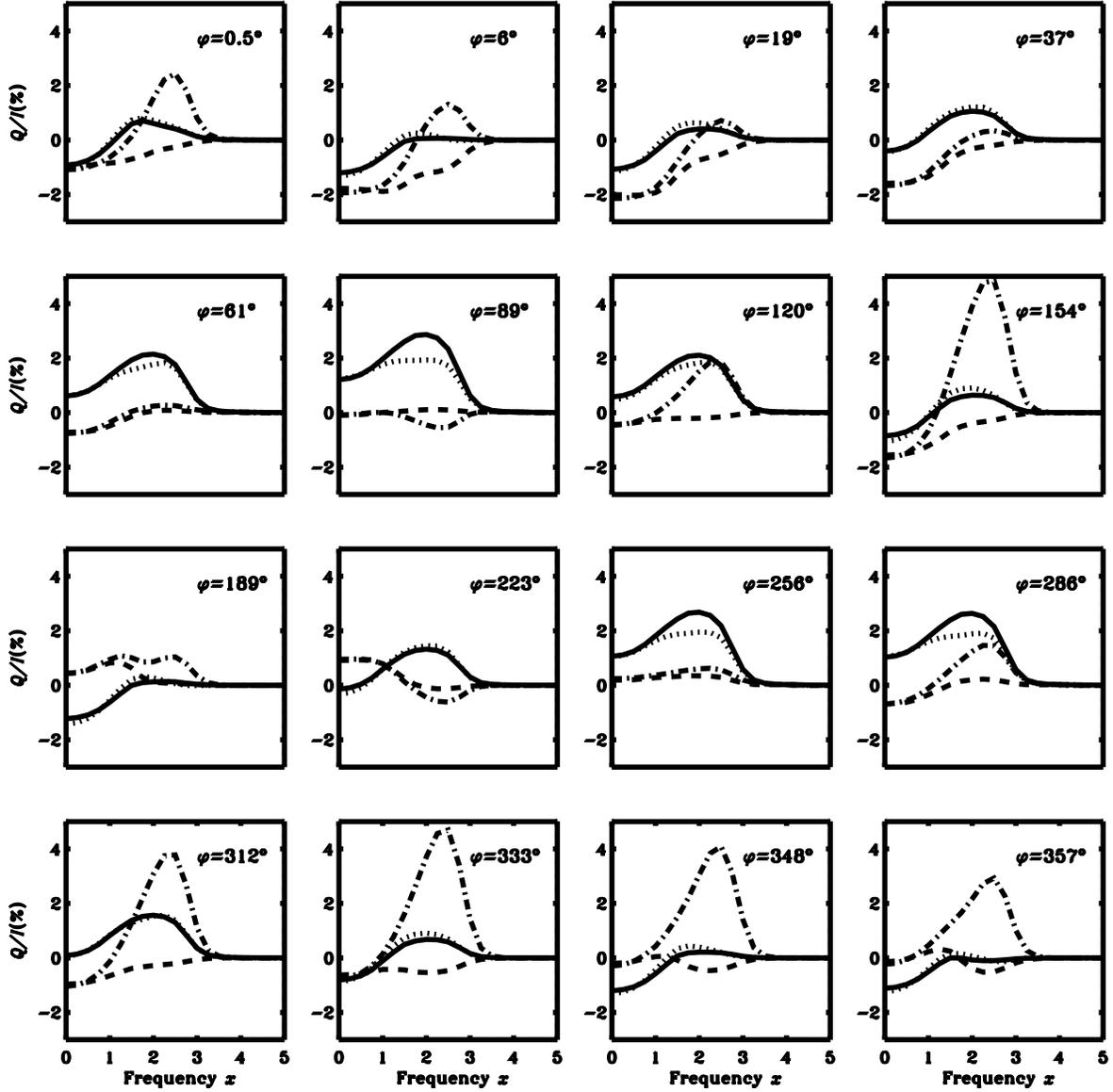}
\caption{Emergent, spatially averaged  $Q/I$ profiles for a 2D medium
with $T_Y=T_Z=20$, for a line of sight $\mu=0.11$. Different panels
correspond to different values of $\varphi$ marked in the panels. 
Solid and dotted lines 
correspond to the AA and the AD profiles for $\bm{B}=0$. 
Dashed and dot-dashed lines correspond to the AA and
the AD profiles in a magnetic medium with magnetic field parameter
$(\Gamma,\theta_B,\chi_B)=(1,90^{\circ},60^{\circ})$.
}
\label{fig-varphi-Q}
\end{figure*}

In Figures~\ref{fig-varphi-Q} and \ref{fig-varphi-U} we present
the emergent, spatially averaged $Q/I$ and $U/I$ profiles
computed using the AD and the AA PRD in line 
scattering for non-magnetic and magnetic 2D media. We show the
results for $\mu=0.11$ and sixteen different values of $\varphi$ (marked on
the respective panels). For the optically thin cases
considered in this paper the AD PRD effects
are restricted to the frequency domain $0 \le x \le5$.
To understand these results
let us consider two examples ($\varphi=0.5^{\circ}$
and 89$^{\circ}$). For $\varphi=0.5^{\circ}$ we can approximate
the emergent $Q$ and $U$ using Equations~(\ref{transform-2}) 
and (\ref{transform-3}) as
\begin{eqnarray}
&&Q(\mu=0.11,\varphi=0.5^{\circ},x)\approx \nonumber \\
&&-\frac{3}{2\sqrt{2}} I^2_0 -\frac{\sqrt{3}}{2} I^{2, \rm x}_2, 
\label{star1}
\end{eqnarray}
and
\begin{eqnarray}
&&U(\mu=0.11,\varphi=0.5^{\circ},x)\approx
{\sqrt{3}}\, I^{2, \rm y}_1.
\label{star2}
\end{eqnarray}

For $\varphi$=89$^{\circ}$ also we can obtain approximate expressions
for $Q$ and $U$  given by
\begin{eqnarray}
&&Q(\mu=0.11,\varphi=89^{\circ},x)\approx \nonumber \\
&&-\frac{3}{2\sqrt{2}} I^2_0+\frac{\sqrt{3}}{2} I^{2, \rm x}_2,
\label{star3}
\end{eqnarray}
and
\begin{eqnarray}
&&U(\mu=0.11,\varphi=89^{\circ},x)\approx
{\sqrt{3}}\, I^{2, \rm x}_1.
\label{star4}
\end{eqnarray}

\subsubsection{Angle-dependent PRD effects in the non-magnetic case}
In both the Figures~\ref{fig-varphi-Q} and \ref{fig-varphi-U}, 
the solid and dotted curves represent the $\bm{B}=0$
case. It is easy to observe that the differences between these curves
depend on the choice of the azimuth angles $\varphi$ for $Q/I$,
while for $U/I$ the differences are marginal.

\subsubsubsection{The $Q/I$ profiles}
For $\varphi=0.5^{\circ}$ the $[Q/I]_{\rm AD}$ and 
$[Q/I]_{\rm AA}$ nearly coincide. But for $\varphi=89^{\circ}$
they differ by $\sim 1\%$ (in the degree of linear polarization) 
around $x=2$, which is very significant. From Equations~(\ref{star1})
and (\ref{star3}) it is clear that $[Q/I]_{\rm AD}$ and
$[Q/I]_{\rm AA}$ are controlled by the combinations of the components
$I^2_0$ and $I^{2,\rm x}_2$. We can see from the first two columns of 
Figure~\ref{fig-I-1to6-nonmag} that 
for $\varphi=0.5^{\circ}$, $I^2_0$ and $I^{2,\rm x}_2$ have comparable 
magnitudes for both the AA and the AD PRD. Further, $[I^2_0]_{\rm AA}<0$,
$[I^{2,\rm x}_2]_{\rm AA}>0$, $[I^2_0]_{\rm AD}>0$ and 
$[I^{2,\rm x}_2]_{\rm AD}<0$. From Equation~(\ref{star1}) 
we can see that in spite of their opposite signs, because of their
comparable magnitudes, the combinations of $I^2_0$ and $I^{2,\rm x}_2$
result in nearly same values of $[Q/I]_{\rm AD}$ and $[Q/I]_{\rm AA}$.
When $\varphi=89^{\circ}$ the components $[I^2_0]_{\rm AA}$, 
$[I^2_0]_{\rm AD}$, $[I^{2,\rm x}_2]_{\rm AA}$ and 
$[I^{2,\rm x}_2]_{\rm AD}$ are of comparable magnitudes. 
Whereas $[I^2_0]_{\rm AA}$
and $[I^{2,\rm x}_2]_{\rm AA}$ have opposite signs, $[I^2_0]_{\rm AD}$
and $[I^{2,\rm x}_2]_{\rm AD}$ have the same sign. Therefore from 
Equation~(\ref{star3}) we see that
$[Q/I]_{\rm AD}$ differs from $[Q/I]_{\rm AA}$ for $\varphi=89^{\circ}$. 

To understand the behaviors of
the components of $I^2_0$ and $I^{2,\rm x}_2$ discussed above,
we can refer to Figures~\ref{fig-I-1to6-nonmag-fourier}, 
\ref{fig-I-1to6-mag-fourier} and 
Table~\ref{table_1}. 
The component $\tilde{I}^{2(0)}_0$ contributes dominantly to $I^2_0$,
and is almost identical to $I^2_0$ because the
contribution from $\tilde{I}^{2(k)}_0$ with $k=1, 2, 3, 4$ are
negligible (for both the values of $\varphi$). When 
$\varphi=0.5^{\circ}$, apart from $\tilde{I}^{2,\rm x\,(0)}_2$, 
the component ${\mathcal Re}\left[\tilde{I}^{2,\rm x\,(2)}_2\right]$ 
makes a significant contribution to $I^{2,\rm x}_2$ and 
$\tilde{I}^{2,\rm x\,(k)}_2$ 
with other values of $k$ vanish (graphically). 
${\mathcal Re}\left[\tilde{I}^{2,\rm x\,(2)}_2\right]$ makes nearly equal 
and opposite contribution as $\tilde{I}^{2,\rm x\,(0)}_2$ when 
$\varphi=0.5^{\circ}$.
When $\varphi=89^{\circ}$,
the contribution of $\tilde{I}^{2,\rm x\,(0)}_2$ is larger than
that of ${\mathcal Re}\left[\tilde{I}^{2,\rm x\,(2)}_2\right]$. 
Also, 
the components $\tilde{I}^{2\,(0)}_0$ and $\tilde{I}^{2,\rm x\,(0)}_2$ 
have the same sign for both the values of $\varphi$. 
Therefore From Equations~(\ref{transform-2f1}) and (\ref{transform-2f2}) 
we can see that $I^2_0$ and $I^{2,\rm x}_2$ have opposite signs for 
$\varphi=0.5^{\circ}$ but have the same signs for $\varphi=89^{\circ}$.

The AD and the AA values of $Q/I$ sometimes
coincide well and sometimes differ significantly. 
This is because, the Fourier components of the AD PRD functions 
$\tilde{r}^{(k)}_{\rm II,III}$ with $k=0$ essentially
represent the azimuthal averages of the AD $r_{\rm II,III}$ functions and are
not same as the explicit angle-averages of the AD $r_{\rm II,III}$ functions.
The latter are obtained by averaging over both co-latitudes and azimuths
(i.e., over all the scattering angles). 
The $\mu$-dependence of the AD  $r_{\rm II,III}$ functions
are contained dominantly in the $\tilde{r}^{(0)}_{\rm II,III}$ 
terms and the $\varphi$-dependence
is contained dominantly in the higher order terms in the Fourier 
expansions of the AD $r_{\rm II,III}$ functions. For this reason 
the AA PRD cannot always be a good representation
of the AD PRD, especially in the 2D polarized line transfer. 
This can be attributed to the strong dependence of the radiation field on the 
azimuth angle ($\varphi$) in the 2D geometry. As will be shown below, the 
differences between the AD and the AA solutions get further 
enhanced in the magnetic case (Hanle effect).

\subsubsubsection{The $U/I$ profiles}
When $\bm{B}=0$, $[U/I]_{\rm AD}$ and $[U/I]_{\rm AA}$ profiles
for both values of $\varphi$ (0.5$^{\circ}$ and 89$^{\circ}$)
do not differ significantly. Equations~(\ref{star2}) and (\ref{star4})
suggest that $U$ has dominant contribution
from $I^{2,{\rm y}}_1$ for $\varphi$=0.5$^{\circ}$
and $I^{2,{\rm x}}_1$ for 89$^{\circ}$. 
Looking at the first two columns of Figure~\ref{fig-I-1to6-nonmag-fourier}, 
it can be seen that $\tilde{I}^{2,\rm y\,(0)}_1$ 
nearly coincide with $[I^{2,\rm y}_1]_{\rm AA}$ for $\varphi=0.5^{\circ}$.
Except $\tilde{I}^{2,\rm y\,(0)}_1$, $\tilde{I}^{2,\rm y\,(k)}_1$
for $k\ne0$ make smaller contribution in the construction of 
$[I^{2,\rm y}_1]_{\rm AD}$. Thus $[I^{2,{\rm y}}_1]_{\rm AA}$ and 
$[I^{2,{\rm y}}_1]_{\rm AD}$ nearly coincide for $\varphi$=0.5$^{\circ}$ 
(see the first two columns of Figure~\ref{fig-I-1to6-nonmag}).
Thus $[U/I]_{\rm AD}$ and $[U/I]_{\rm AA}$ are nearly
the same for $\varphi$=0.5$^{\circ}$.
When  $\varphi$=89$^{\circ}$ (the first two columns of 
Figure~\ref{fig-I-1to6-mag}), $[I^{2,{\rm x}}_1]_{\rm AA}$ vanishes.
For each $k$, $\tilde{I}^{2,\rm x(k)}_1$ approach zero, as does
$[I^{2,\rm x}_1]_{\rm AD}$, which is a combination of
$\tilde{I}^{2,\rm x(k)}_1$.
Thus $[U/I]_{\rm AD}$ and $[U/I]_{\rm AA}$ both are nearly
zero for $\varphi$=89$^{\circ}$.
We can carry out similar analysis and find out which are the
irreducible Fourier components of $\tilde{\bm{\mathcal I}}^{(k)}$ 
that contribute to the construction of $\bm{\mathcal I}$
and which of the components of $\bm{\mathcal {I}}$ contribute
to generate $Q$ and $U$ to interpret their behaviors.

\subsubsection{Angle-dependent PRD effects in the 
magnetic case}
\begin{figure*}
\centering
\includegraphics[scale=0.7]{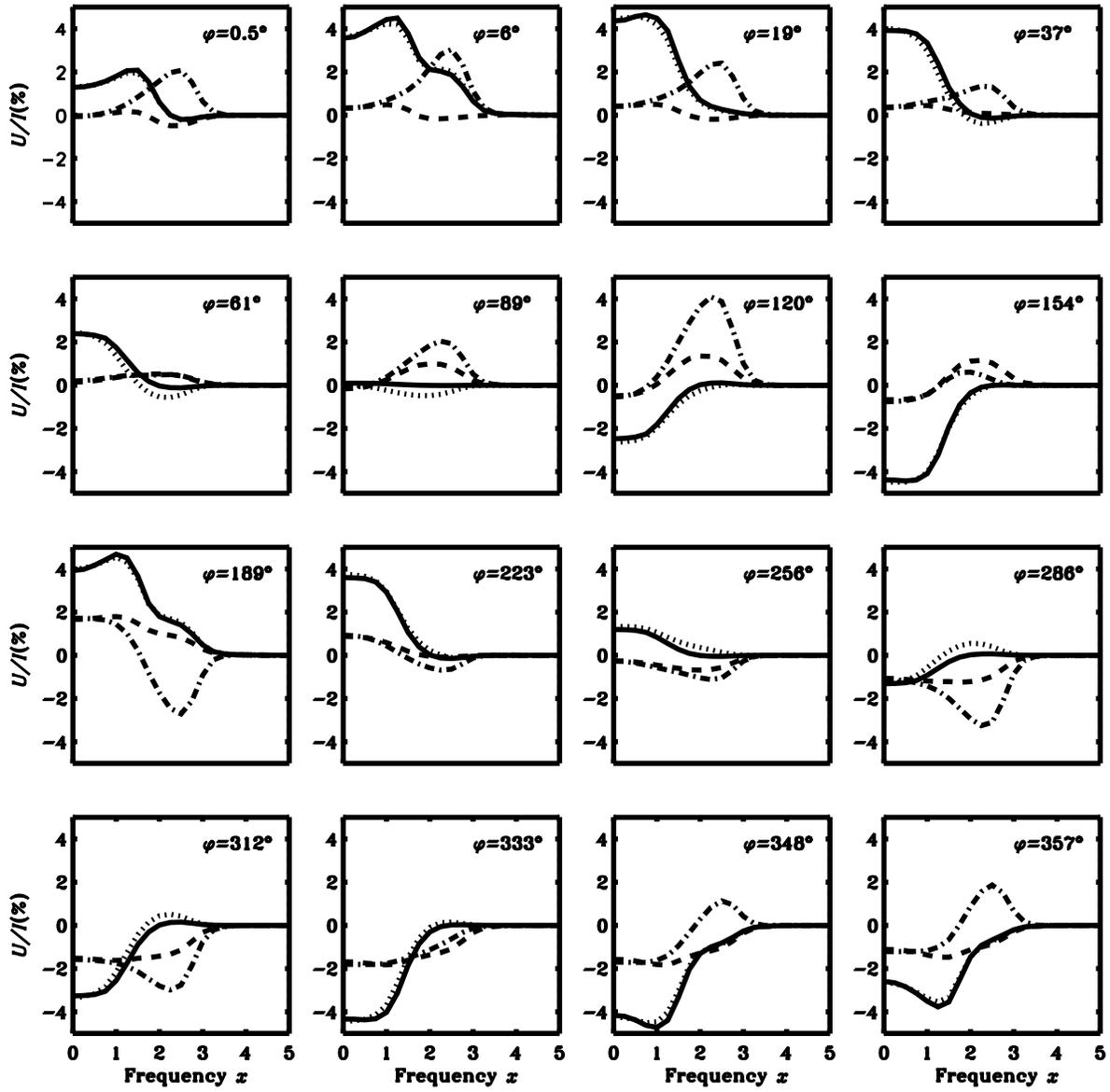}
\caption{Same as Figure~\ref{fig-varphi-Q} but for $U/I$.
}
\label{fig-varphi-U}
\end{figure*}

The presence of a weak, 
oriented magnetic field modifies the values of $Q/I$ and $U/I$ in the 
line core ($x \le 3.5$) to a considerable extent, due to Hanle effect.
Further, it is for $\bm{B}\ne0$ that the differences between the AA 
and the AD PRD become more significant. 
In both the Figures~\ref{fig-varphi-Q} and \ref{fig-varphi-U},
the dashed and dot-dashed curves represent $\bm{B}\ne0$ case.
As usual, there is either a depolarization (decrease in the magnitude)
or a re-polarization (increase in the magnitude) of both $Q/I$ and $U/I$
with respect to those in the $\bm{B}=0$ case. The AD PRD values of $Q/I$
and $U/I$ are larger in magnitude (absolute values) than those of the 
AA PRD, for the chosen set of model parameters (this is not to be taken as a 
general conclusion). The differences depend sensitively on the value of 
$\bm{B}$.

\begin{figure*}
\centering
\includegraphics[scale=0.4]{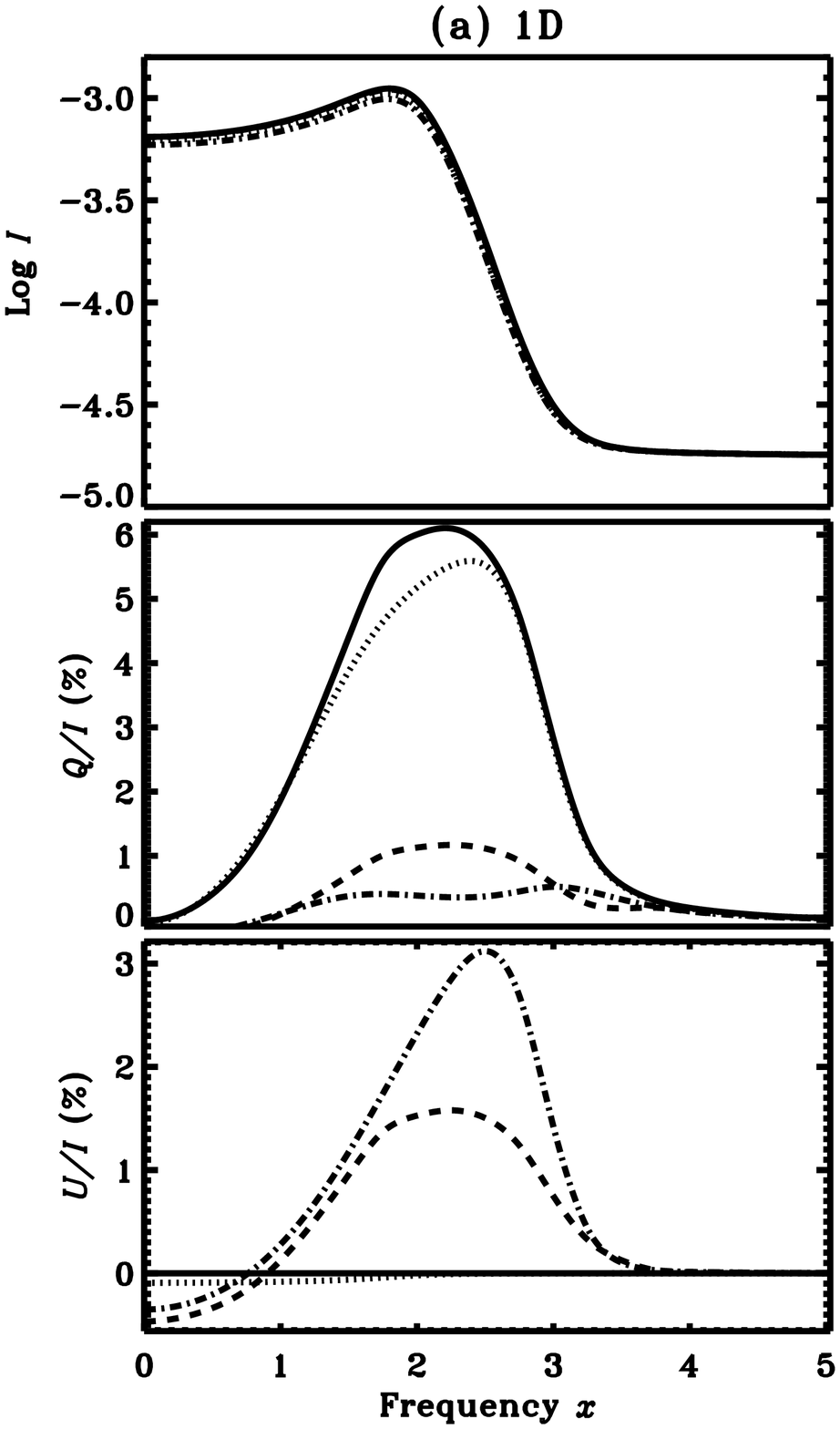}
\hspace{-1cm}
\includegraphics[scale=0.4]{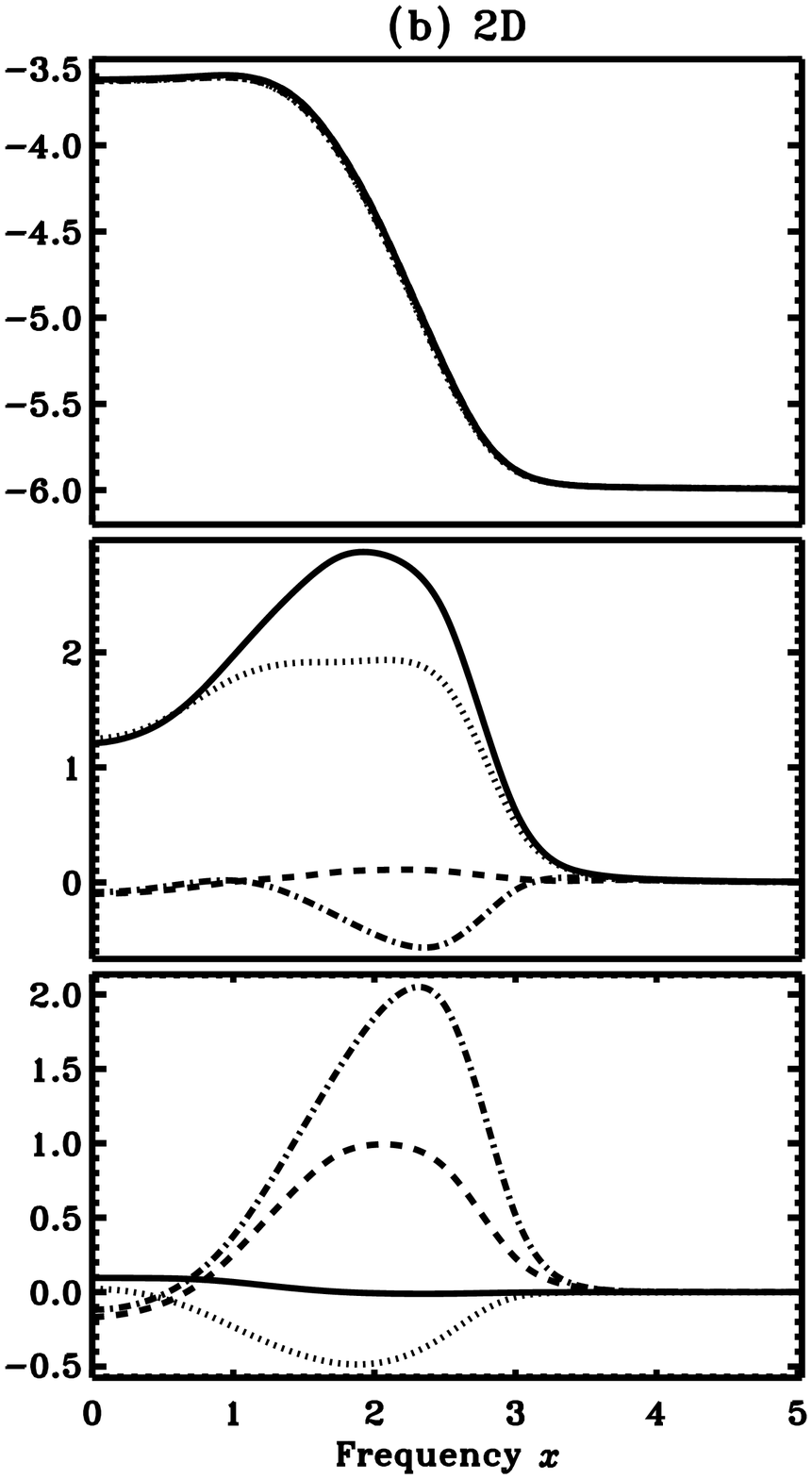}
\caption{Panel (a) shows emergent $(I,Q/I,U/I)$ profiles formed in an 1D 
medium and the panel (b) shows the emergent, spatially averaged $(I,Q/I,U/I)$
profiles formed in a 2D medium. The solid and dotted lines represent 
respectively the AA and the AD profiles for $\bm{B}=0$. 
The dashed and dash-triple-dotted lines represent 
respectively the AA and the AD profiles for $\bm{B}\ne0$, 
with the magnetic field parameterized by 
$(\Gamma,\theta_B,\chi_B)=(1,90^{\circ},60^{\circ})$. The results are
shown for $\mu=0.11$ and $\varphi=89^{\circ}$. For the panel (a)
we take $T_Z=T=20$ and for the panel (b), $T_Z=T_Y=T=20$.
}
\label{fig-2D-T20}
\end{figure*}

\subsubsubsection{Comparison with 1D results}
In Figures~\ref{fig-2D-T20}(a) and (b) we present the emergent $(I,Q/I,U/I)$
profiles for 1D and 2D media for $\mu=0.11$ and $\varphi=89^{\circ}$. 
For 2D RT, we present the spatially 
averaged profiles. 
The effects of a multi-D geometry (2D or 3D) on linear polarization 
for non-magnetic and magnetic cases are 
discussed in detail in Papers I, II and III, where we considered 
polarized line formation in multi-D media, scattering according to 
the AA PRD. We recall here that the essential
effects are due to the finite boundaries in multi-D media, which cause
leaking of radiation and hence a decrease in the values of Stokes $I$, 
and a sharp rise in the values of $Q/I$ and $U/I$ near the boundaries. 
Multi-D geometry naturally breaks the
axisymmetry of the medium that prevails in a 1D planar medium. This leads
to significant differences in the values of $Q/I$ and $U/I$ formed
in 1D and multi-D media (compare solid lines in panels (a) and (b) of 
Figure~ \ref{fig-2D-T20}). 
As pointed out in Papers I, II and III, for non-magnetic case, 
$U/I$ is zero in 1D media while in 2D media a non-zero $U/I$
is generated due to symmetry breaking by the finite boundaries.
For the $(\theta,\varphi)$ values chosen in Figure~\ref{fig-2D-T20}(b)
$[U/I]_{\rm AA}$is nearly zero even for non-magnetic 2D case, which is
not generally true for other choices of $(\theta,\varphi)$ (see solid 
lines in various panels of Figure~\ref{fig-varphi-U}). 

The effects of the AD PRD in $Q/I$ and $U/I$ profiles are already 
discussed above for non-magnetic and magnetic 2D media.
They are similar for both 1D and
2D cases. For the non-magnetic 2D media, we can see the AD PRD effects 
even in $U/I$, which is absent in the corresponding 1D media. 
In 1D, one has to apply a non-zero magnetic
field $\bm{B}$ in order to see the effects of the AD PRD on $U/I$ profiles. 

The magnitudes of $[Q/I]_{\rm 1D}$ in the non-magnetic case and
of $[Q/I]_{\rm 1D}$, $[U/I]_{\rm 1D}$ in the magnetic case are larger
in comparison with the corresponding spatially averaged $[Q/I]_{\rm 2D}$ 
and $[U/I]_{\rm 2D}$. This is again due to leaking of photons from the 
finite boundaries and the effect of spatial averaging (which causes 
cancellation of positive and negative quantities).

\subsection{Radiation anisotropy in 2D media--Stokes source vectors}
\begin{figure*}
\centering
\includegraphics[scale=0.25]{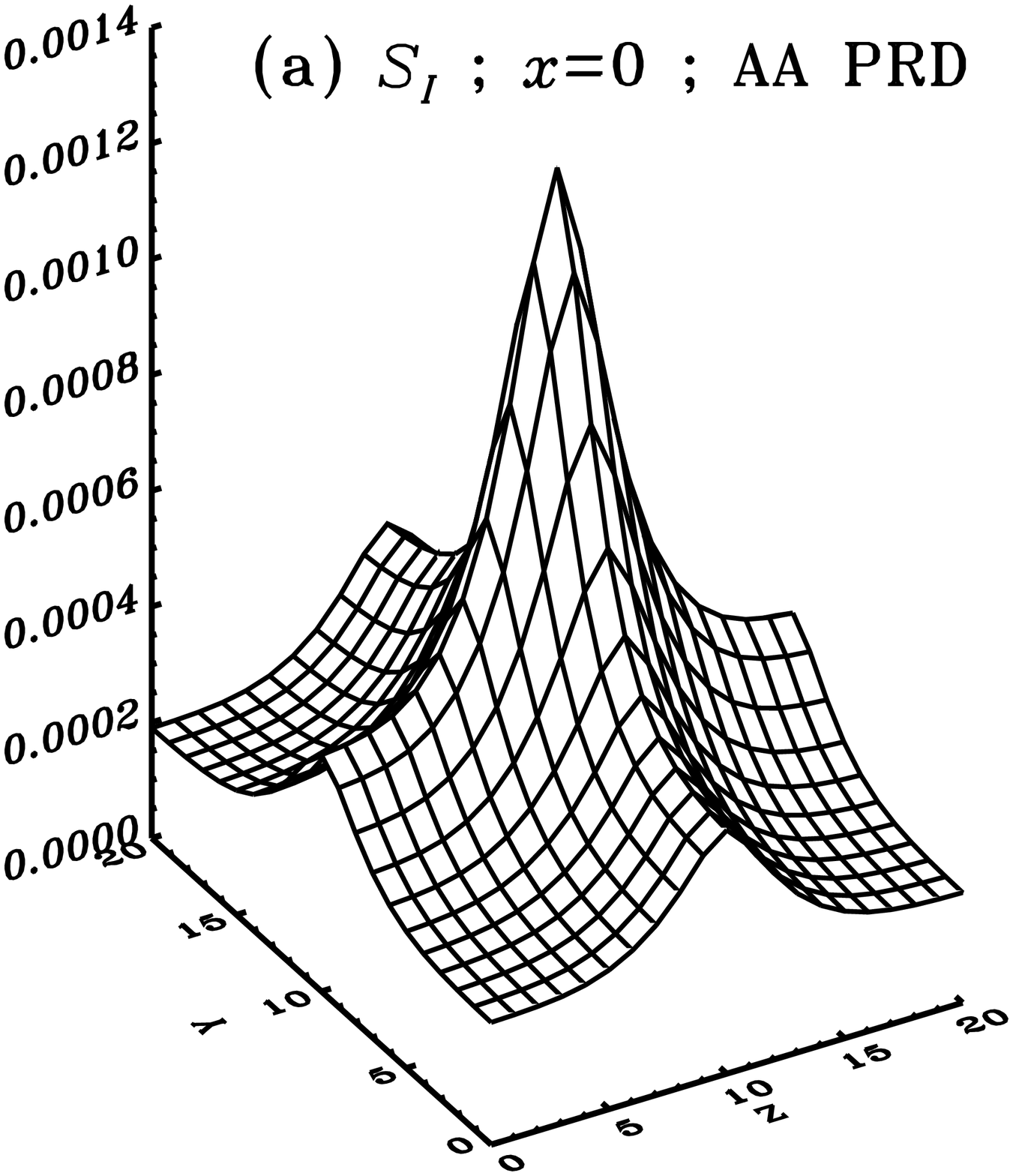}
\includegraphics[scale=0.25]{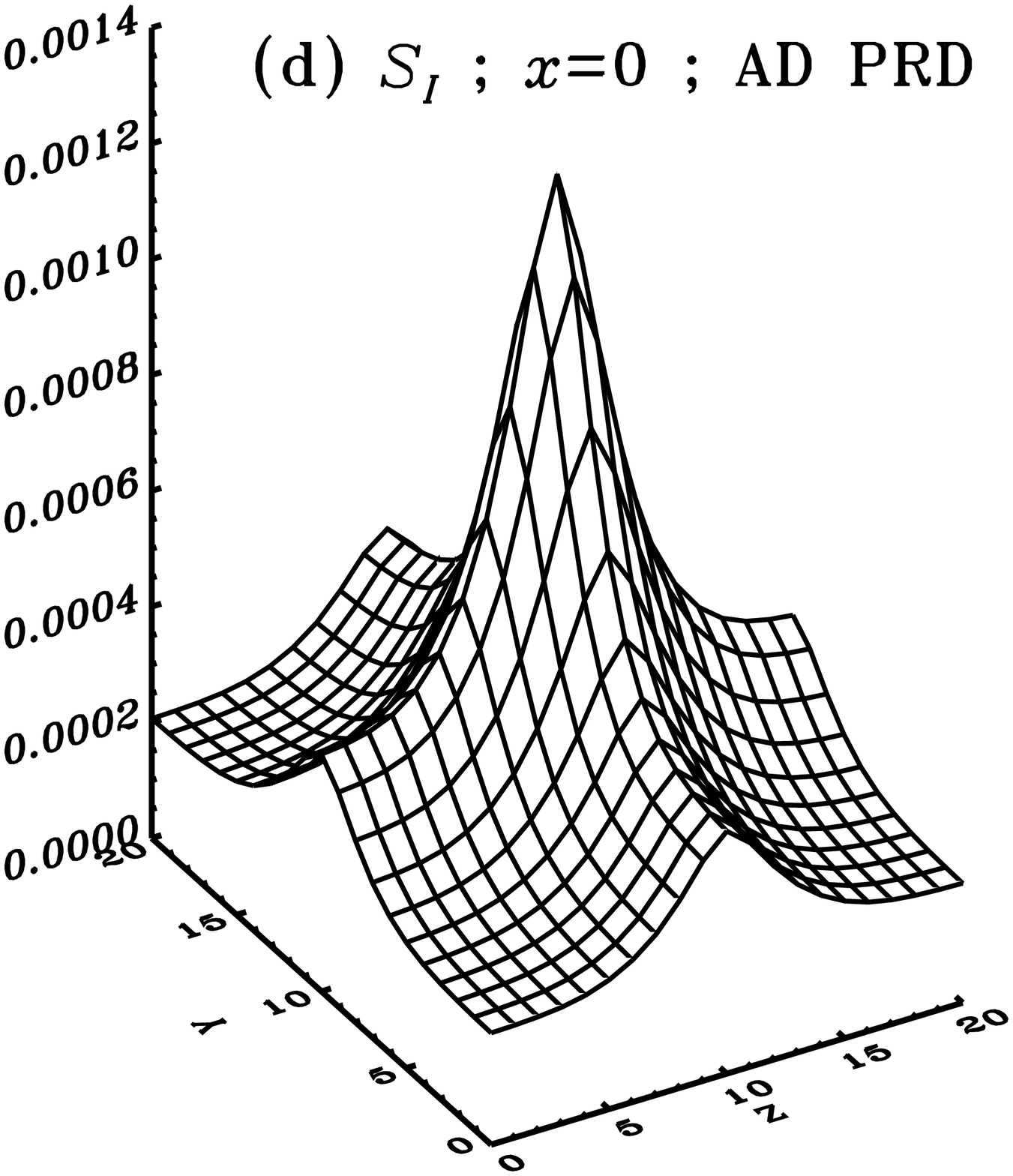}
\includegraphics[scale=0.25]{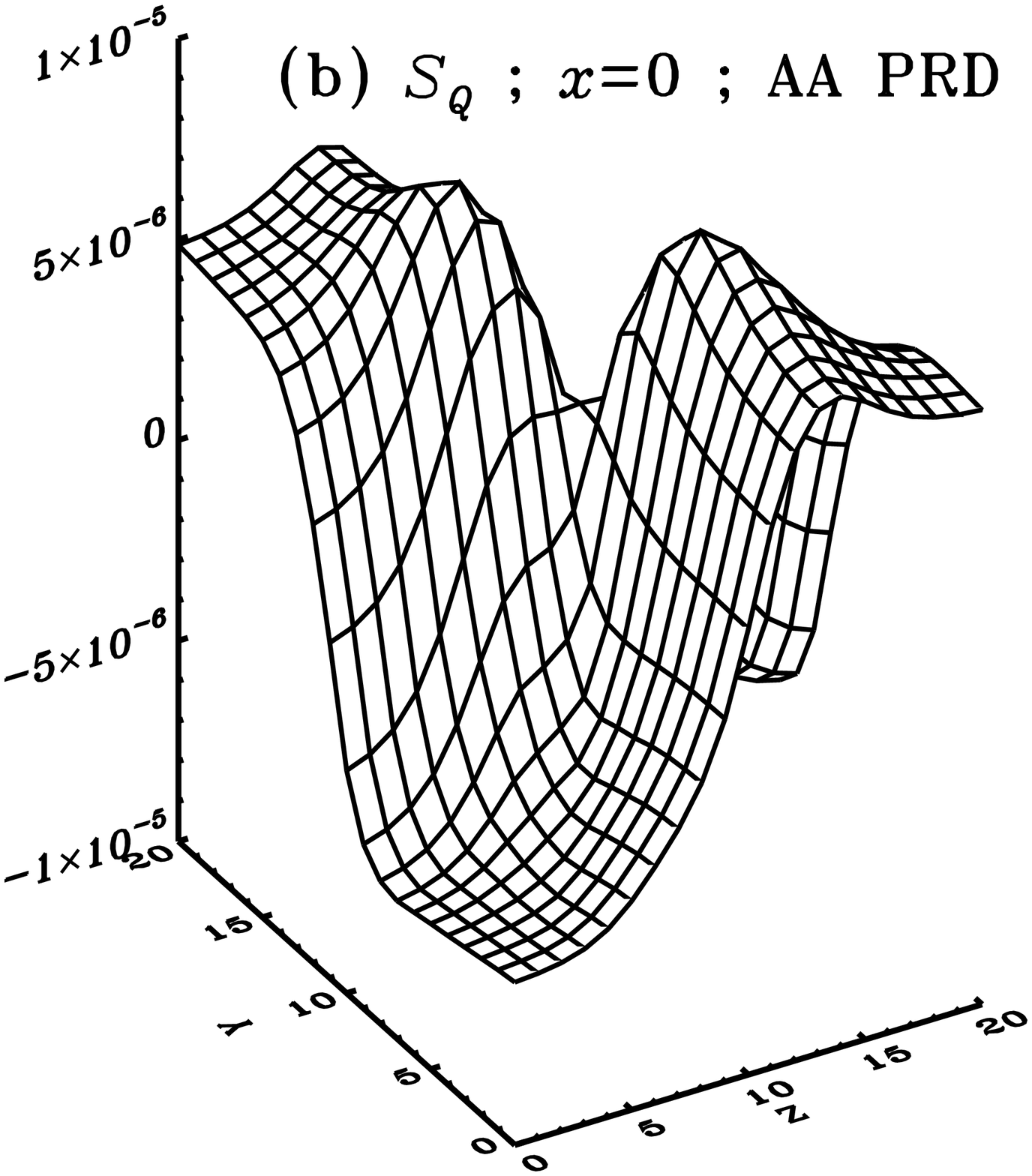}
\includegraphics[scale=0.25]{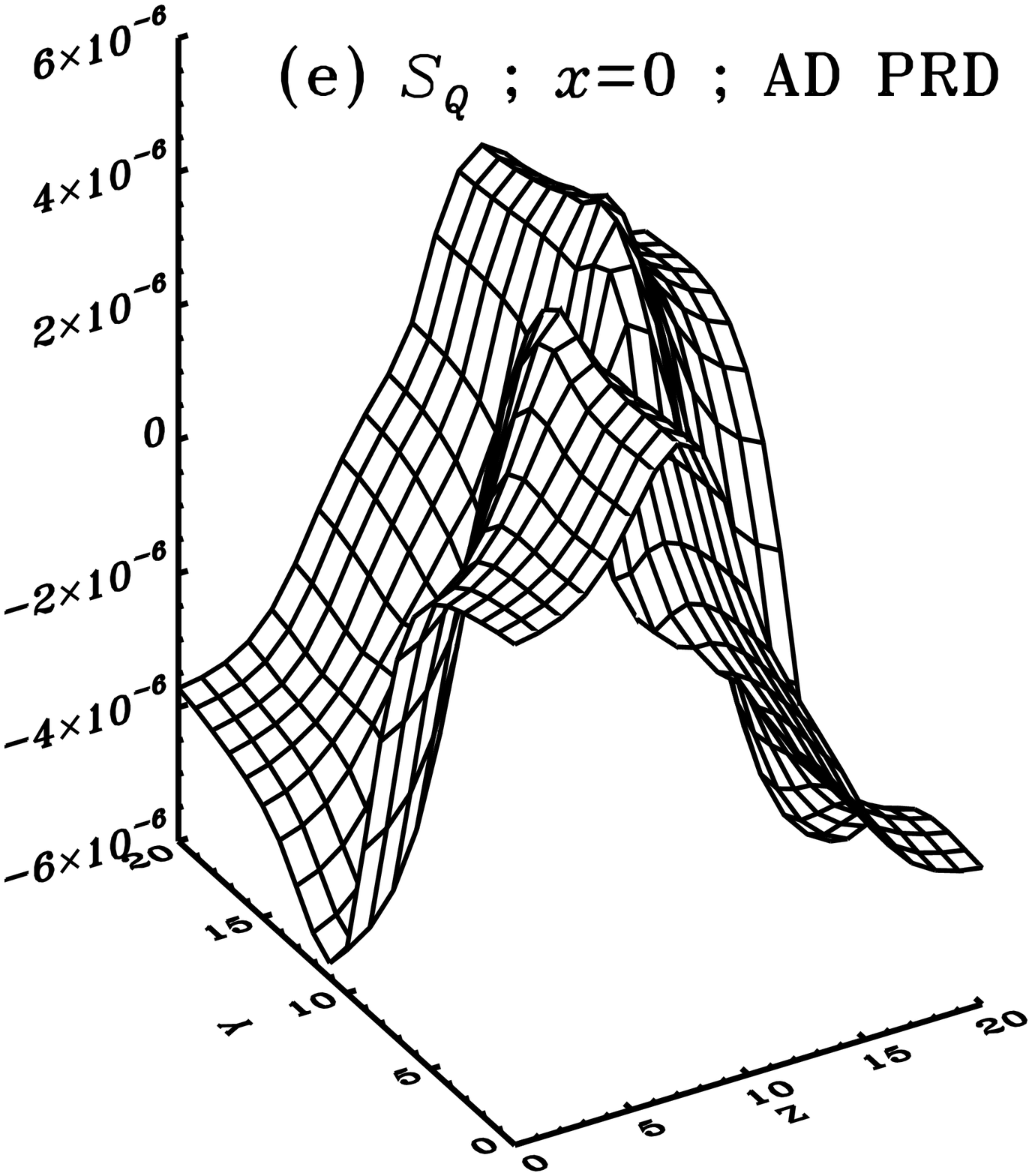}
\includegraphics[scale=0.25]{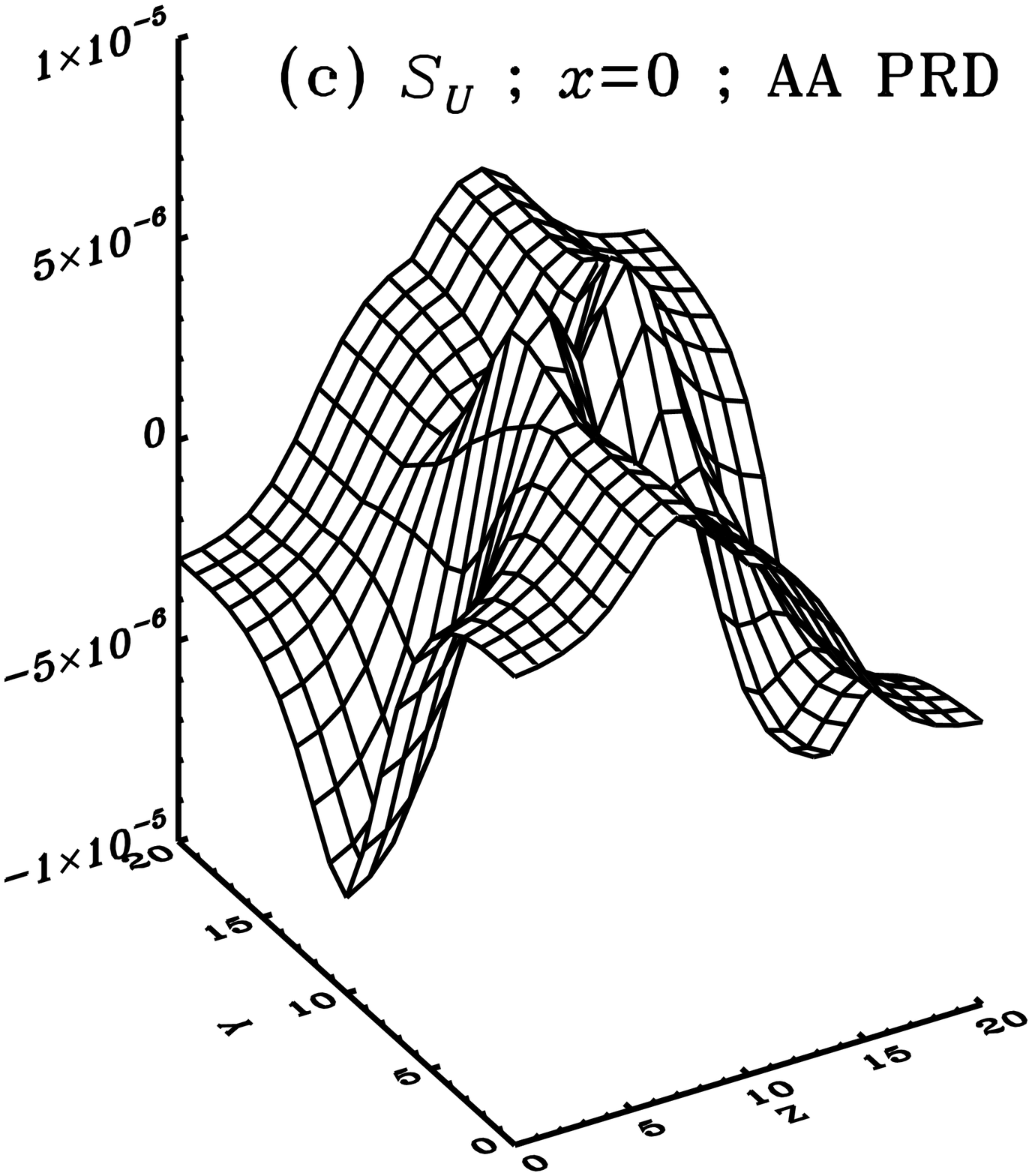}
\includegraphics[scale=0.25]{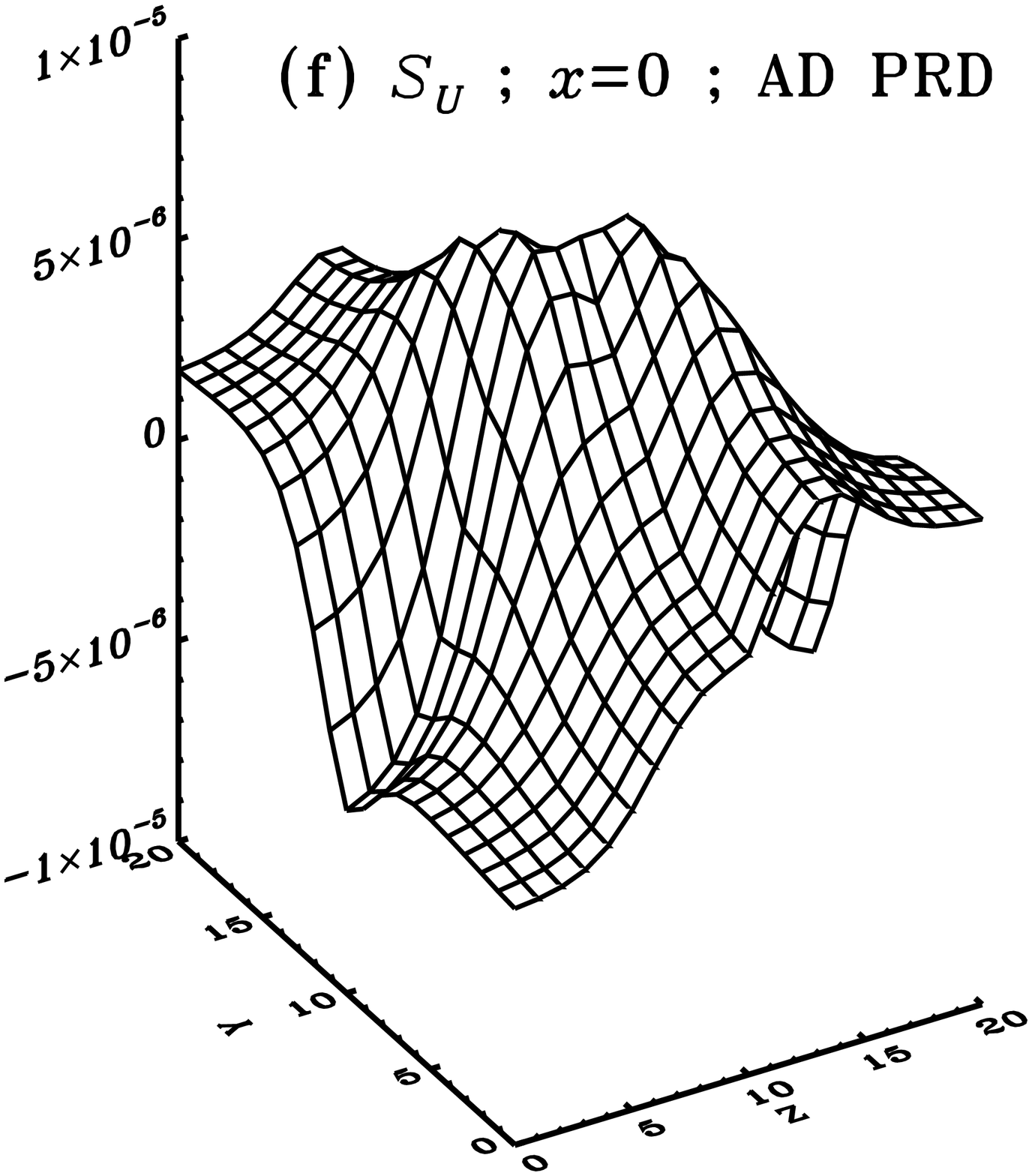}
\caption{Surface plots of $S_I$, $S_Q$ and $S_U$ for the AA
(left panels) and the AD PRD (right panels) for $x=0$. The source vector
components are plotted as a function of the grid indices along $Y$ and 
$Z$ directions. Here $\bm{B}\ne 0$, with 
$(\Gamma,\theta_B,\chi_B)=(1,90^{\circ},60^{\circ})$.
The other model parameters are same as
in Figure~\ref{fig-2D-T20}. }
\label{fig-2D-surface-AA-AD-x0}
\end{figure*}

\begin{figure*}
\centering
\includegraphics[scale=0.25]{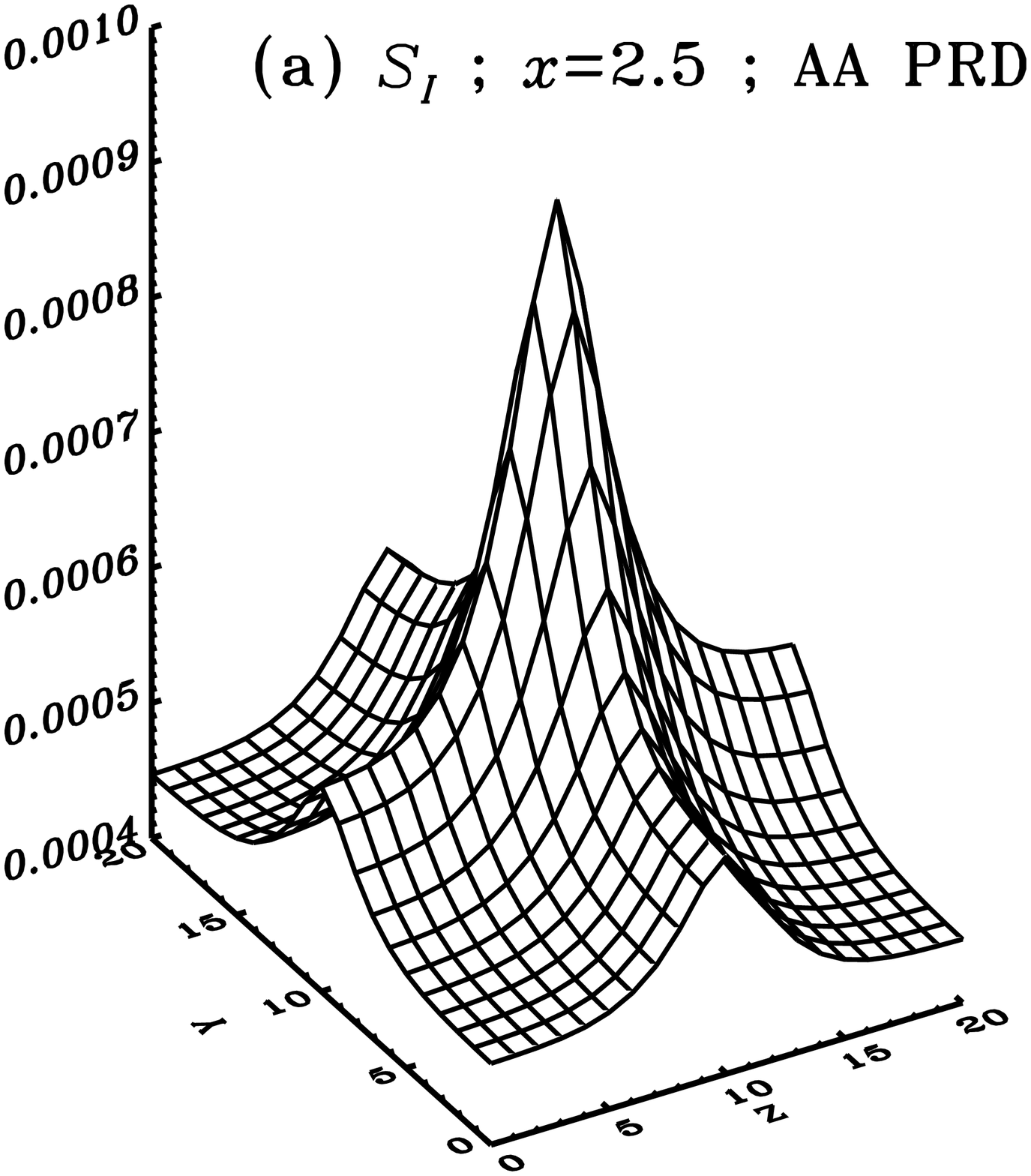}
\includegraphics[scale=0.25]{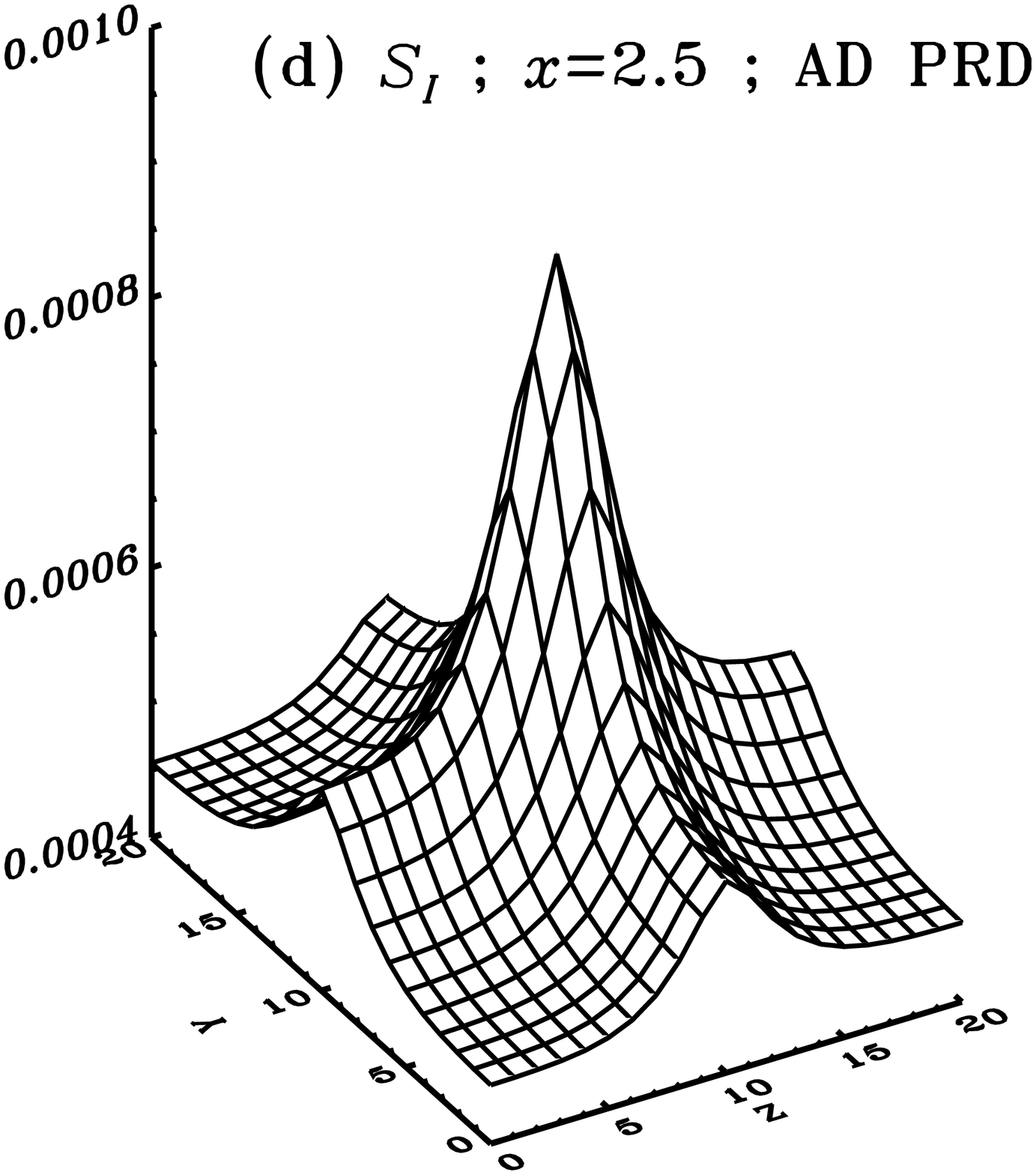}
\includegraphics[scale=0.25]{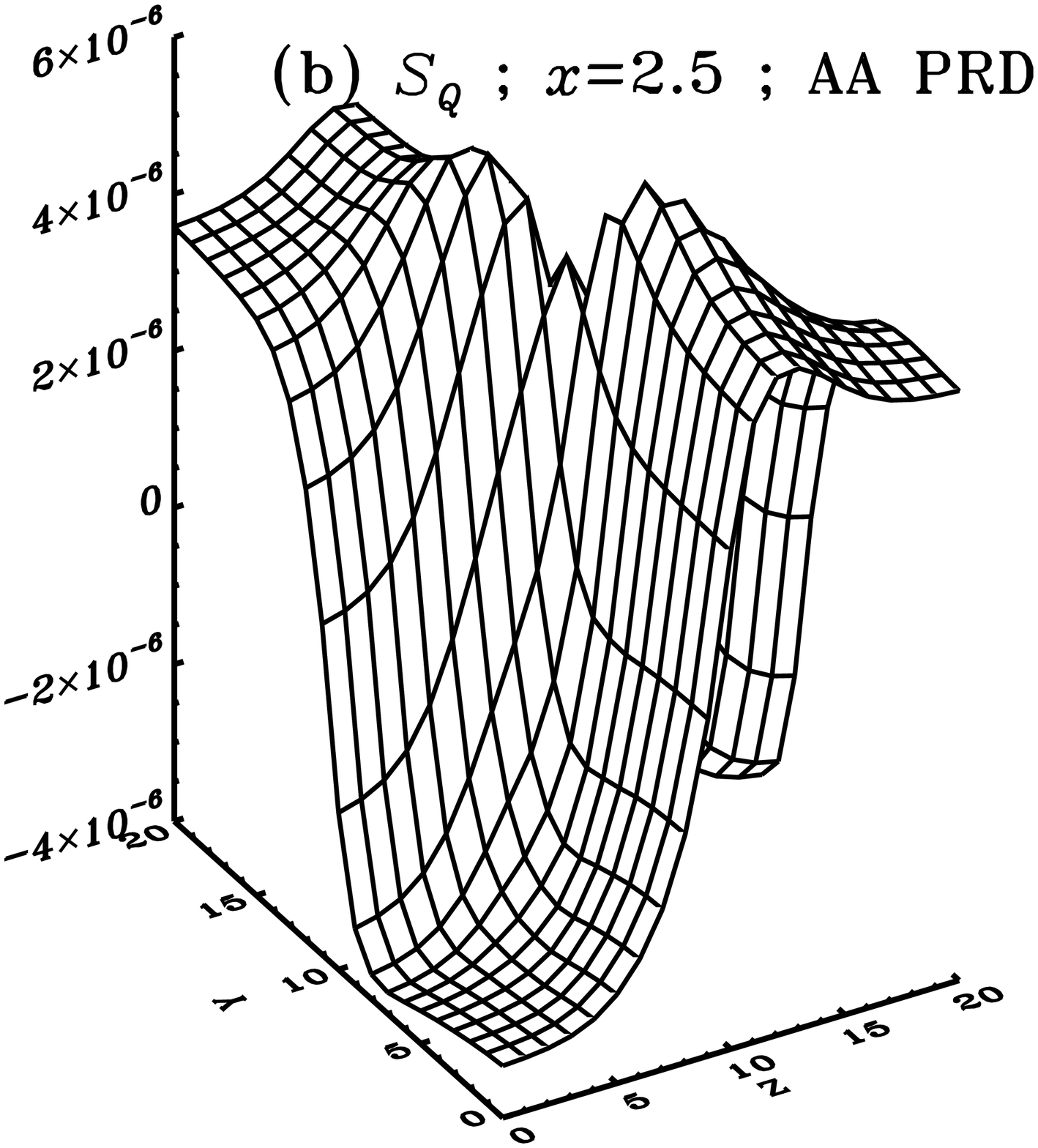}
\includegraphics[scale=0.25]{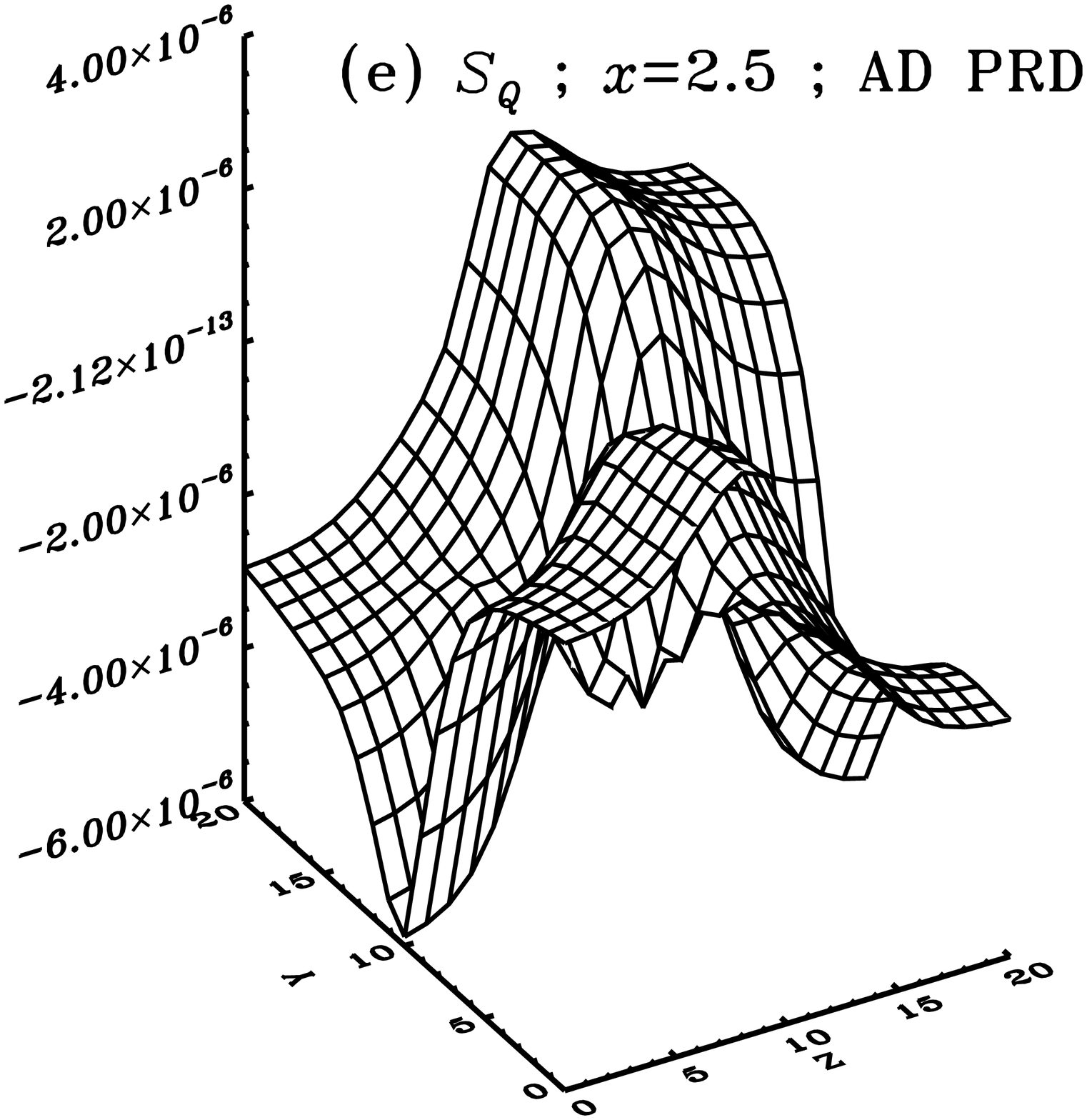}
\includegraphics[scale=0.25]{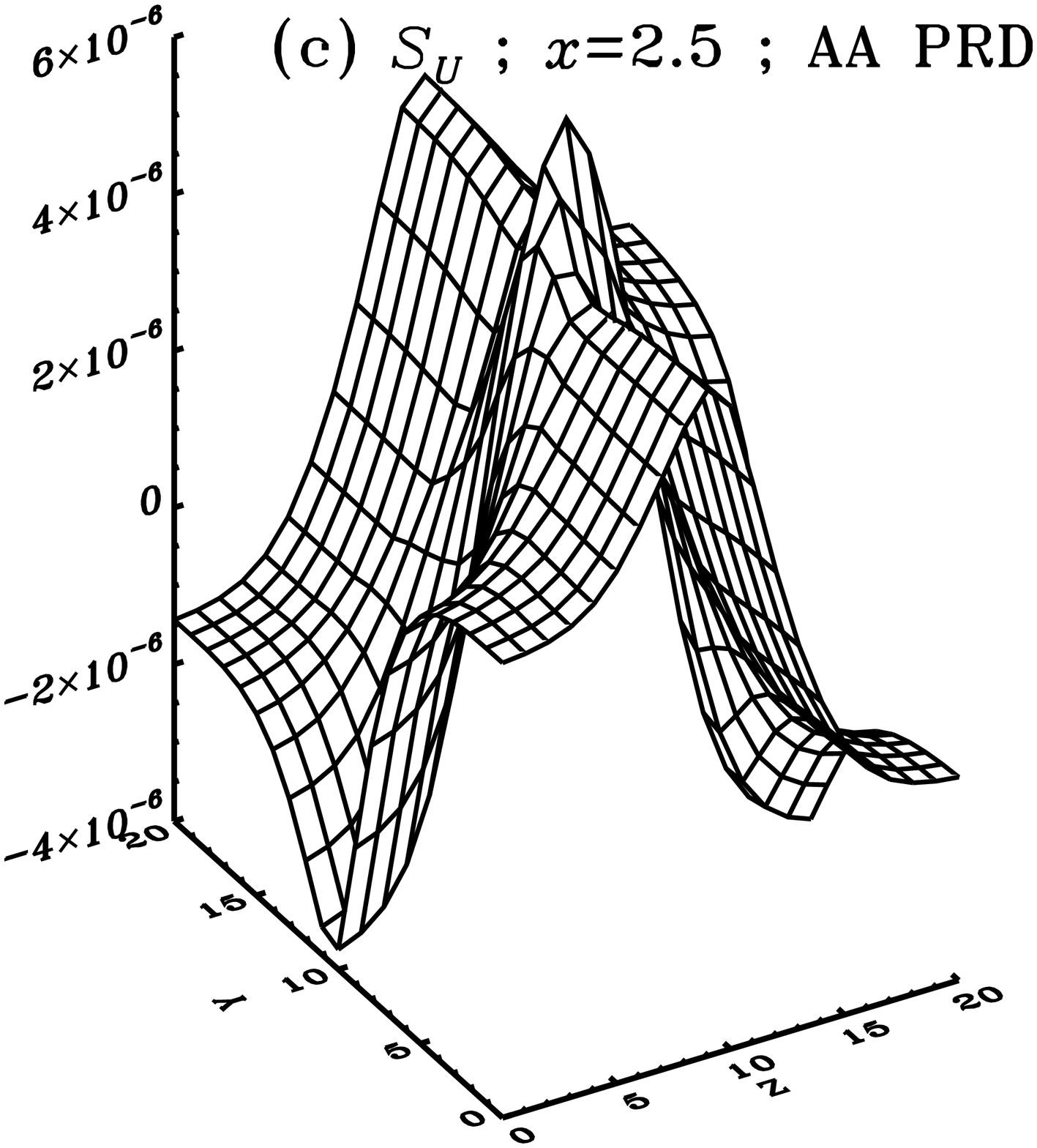}
\includegraphics[scale=0.25]{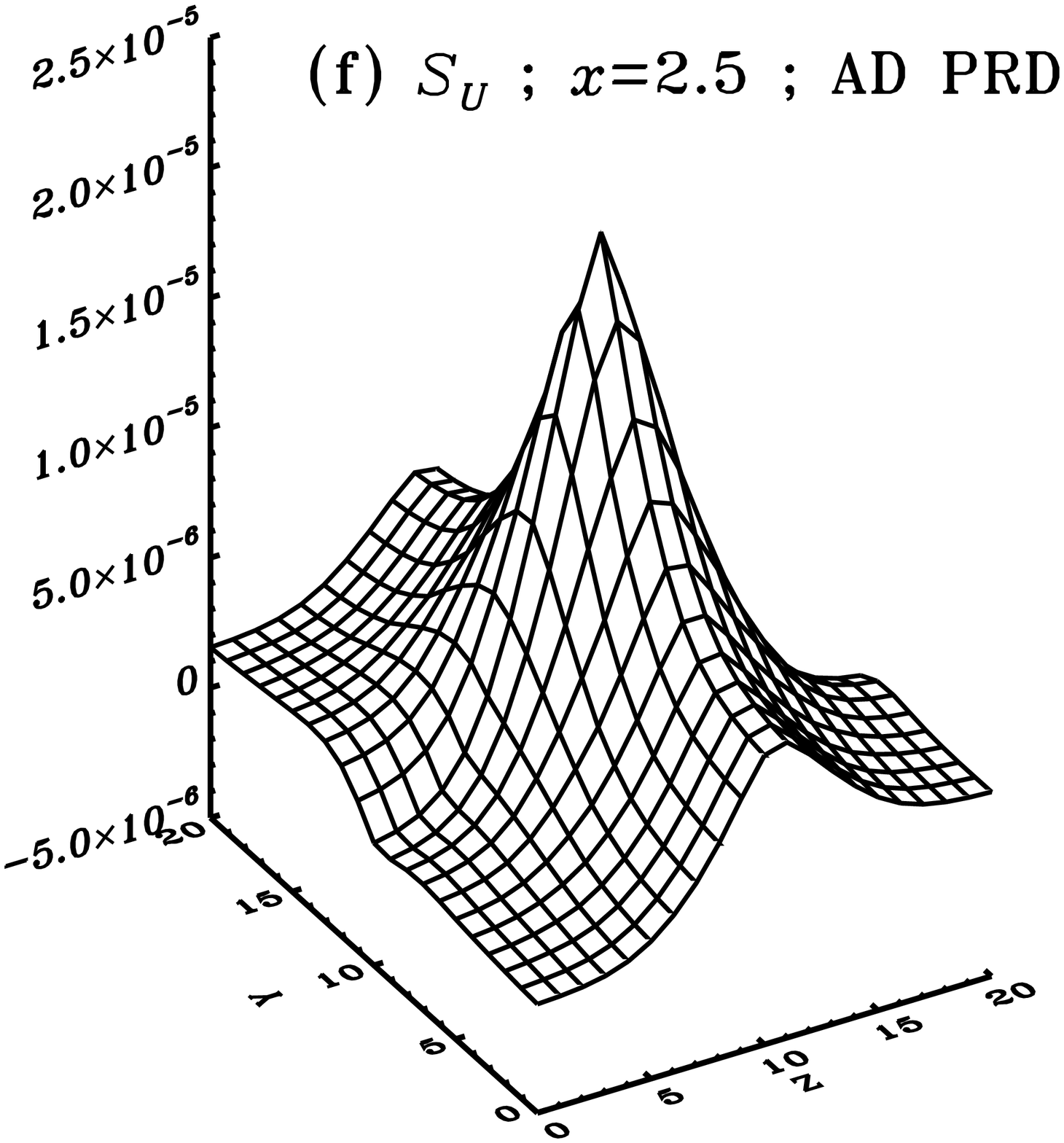}
\caption{Same as Figure~\ref{fig-2D-surface-AA-AD-x0}
for $x=2.5$.}
\label{fig-2D-surface-AA-AD-x2.5}
\end{figure*}

In Figures~\ref{fig-2D-surface-AA-AD-x0} and \ref{fig-2D-surface-AA-AD-x2.5}
we present spatial distribution of $S_I$, $S_Q$ and $S_U$ on the plane of the 
2D slab for two different frequencies ($x=0$ and $x=2.5$ respectively).
The spatial distribution of source vector components $S_Q$ and $S_U$
represent the anisotropy of the radiation field in the 2D medium. 
It shows how inhomogeneous is the distribution of linear polarization
within the 2D medium. 

In Figure~\ref{fig-2D-surface-AA-AD-x0} we consider $x=0$ (line center).
For the chosen values of $(\theta,\varphi)$ the spatial distribution
of $S_I$ is not very different for the AA and the AD PRD. $S_Q$
and $S_U$ for both the AA and the AD PRD have similar magnitudes
(Figures~\ref{fig-2D-surface-AA-AD-x0}(b),(c) and 
\ref{fig-2D-surface-AA-AD-x0}(e),(f)), but different spatial distributions.
The spatial distribution of $S_Q$ and $S_U$ is such that
the positive and negative contributions with similar magnitudes of
$S_Q$ and $S_U$ cancel out in the computation of their formal integrals.
Therefore, the average values of $Q/I$ and $U/I$ resulting from the
formal integrals of $S_Q$ and $S_U$ are nearly zero at $x=0$ for both
the AA and the AD PRD (see dashed and dot-dashed lines at $x=0$ 
in Figure~\ref{fig-2D-T20}(b)).

In Figure~\ref{fig-2D-surface-AA-AD-x2.5} we consider $x=2.5$ (near wing
frequency). Again, $S_I$ does not show significant differences between the AA
and the AD PRD. For $S_Q$, the AA PRD has a distribution with
positive and negative values equally distributed in the 2D slab
but the AD PRD has more negative contribution. This reflects in the average
values of $Q/I$, where $[Q/I]_{\rm AA}$ approach zero due to cancellation, 
while $[Q/I]_{\rm AD}$ values are more negative (see dashed and 
dot-dashed lines at $x=2.5$ in Figure~\ref{fig-2D-T20}(b)).
The positive and negative values of $S_U$ are distributed in a 
complicated manner everywhere on the 2D slab for the AA PRD. 
For the AD PRD, the distribution of $S_U$
is positive almost everywhere, including the central parts of the 2D slab. 
Such a spatial distribution reflects again in the average value of $U/I$ 
(shown in Figure~\ref{fig-2D-T20}(b)), where $[U/I]_{\rm AA}$ have
smaller positive magnitudes (due to cancellation effects) than the 
corresponding $[U/I]_{\rm AD}$.

\section{CONCLUSIONS}
\label{conclusions}
In this paper we have further generalized the Fourier decomposition 
technique developed in Paper IV to handle the AD PRD in multi-D 
polarized RT (see Section~\ref{rteff}). 
We have applied this technique and developed an efficient iterative
method called Pre-BiCG-STAB to solve this problem (see 
Section~\ref{numerics}). 

We prove in this paper that the symmetry of the polarized radiation field
with respect to the infinite axis, 
that exists for a non-magnetic 2D medium for the AA PRD (as shown in
Paper II) breaks down for the AD PRD (see Appendix~\ref{appendixa}).

We present results of the very first 
investigations of the effects of the AD PRD on the polarized line formation 
in multi-D media. We restrict our attention to freestanding 2D slabs
with finite optical thicknesses on the two axes ($Y$ and $Z$). The optical 
thicknesses of the isothermal 2D media considered in this paper are very 
moderate ($T=20$). We consider
effects of the AD PRD on the scattering polarization in both non-magnetic and
magnetic cases. We find that the relative AD PRD effects are prominent
in the magnetic case (Hanle effect). They are also present in non-magnetic 
case for some choices of $(\theta,\varphi)$. We conclude that the AD 
PRD effects
are important for interpreting the observations of scattering polarization
in multi-D structures on the Sun.

Practically, even with the existing advanced computing facilities, it is 
extremely difficult to carryout the multi-D polarized RT with the AD PRD
in spite of using advanced numerical techniques. Therefore in 
this paper we restrict our attention to isothermal 2D slabs. The 
use of the AD PRD in
3D polarized RT in realistic modeling of the observed 
scattering polarization on the Sun will be numerically very expensive 
and can be taken up in future only with highly advanced computing
facilities.

{\bf Erratum}:\,
In the previous papers of this series (Papers I, III and IV) the definitions
of the formal solutions expressed in terms of the optical thicknesses 
have a notational error. In Equation~(20) of Paper I, 
Equations~(14) and (20) of 
Paper III, Equation~(14) of Paper IV, the symbol $\tau_{x,\rm max}$
should have been $\tau_{x}(\bm{r},\bm{\Omega})$ as explicitly
given in Equation~(\ref{i-out-tau}) of this paper. 
$\tau_{x}(\bm{r},\bm{\Omega})$ is defined in Equation~(\ref{tau})
in this paper.
In the previous papers of this series (Papers I to IV) 
the vector $\bm{r}'=\bm{r}-(s-s')
\bm{\Omega}$ was incorrectly defined as $\bm{r}-s'\bm{\Omega}$.
We note here that the numerical
results and all other equations presented in Papers I -- IV are correct,
and are unaffected by this error in the above mentioned equations.

\acknowledgments
We thank the anonymous Referee for very useful comments and suggestions that
helped improve the manuscript to a great extent. The reports by the Referee
helped to correct some of the mistakes that were present in the previous
papers of this series, and the corrections are now presented in the form of 
an erratum in this paper. We also thank the Referee for providing 
Figure~\ref{fig-fs}.

\appendix
\section{SYMMETRY BREAKING PROPERTIES OF THE AD PRD FUNCTIONS
IN NON-MAGNETIC 2D MEDIA}
\label{appendixa}
In this appendix we show that the symmetry properties
that are valid for the AA PRD (proved in Paper II) break down
for the AD PRD.  We present the proof in the form of an algorithm.\\

\noindent
Step (1):\, First we assume that the medium contains only an unpolarized thermal
source namely, $\bm{\mathcal{S}}=(\epsilon B(\bm{r}),0,0,0,0,0)^T$.\\

\noindent
Step (2):\, Use of this source vector in the formal solution expression
yields $\bm{\mathcal{I}}=(I^0_0,0,0,0,0,0)^T$.\\

\noindent
Step 3:\, Using this $\bm{\mathcal{I}}$ we can write the
expressions for the irreducible polarized mean intensity components as
\begin{eqnarray}
&&J^0_0(\bm{r},\bm{\Omega},x)
\simeq \int_{x',\bm{\Omega}'}\frac{\hat{R}(x,x',\bm{\Omega},\bm{\Omega'})}
{\phi(x)} \,I^0_0(\bm{r},\theta',\varphi',x'),
\nonumber \\
&&J^2_0(\bm{r},\bm{\Omega},x)\simeq c_2 \int_{x',\bm{\Omega}'}
\frac{\hat{R}(x,x',\bm{\Omega},\bm{\Omega}')}{\phi(x)}  
(3 \cos^2 \theta' -1)\,I^0_0(\bm{r},\theta',\varphi',x'),
\nonumber \\
&&J^{\rm 2,x}_1(\bm{r},\bm{\Omega},x)\simeq -c_3\int_{x',\bm{\Omega}'}
\frac{\hat{R}(x,x',\bm{\Omega},\bm{\Omega}')}{\phi(x)}  
\sin 2 \theta' \cos \varphi' \,I^0_0(\bm{r},\theta',\varphi',x'),
\nonumber \\
&&J^{\rm 2,y}_1(\bm{r},\bm{\Omega},x)\simeq c_4\int_{x',\bm{\Omega}'}
\frac{\hat{R}(x,x',\bm{\Omega},\bm{\Omega}')}{\phi(x)}  
\sin 2 \theta' \sin \varphi'\,I^0_0(\bm{r},\theta',\varphi',x'),
\nonumber \\
&&J^{\rm 2,x}_2(\bm{r},\bm{\Omega},x)\simeq c_5\int_{x',\bm{\Omega}'}
\frac{\hat{R}(x,x',\bm{\Omega},\bm{\Omega}')}{\phi(x)}  
\sin^2 \theta' \cos 2 \varphi'\,I^0_0(\bm{r},\theta',\varphi',x'),
\nonumber \\
&&J^{\rm 2,y}_2(\bm{r},\bm{\Omega},x)\simeq -c_6\int_{x',\bm{\Omega}'}
\frac{\hat{R}(x,x',\bm{\Omega},\bm{\Omega}')}{\phi(x)}  \sin^2 \theta' 
\sin 2 \varphi' \,I^0_0(\bm{r},\theta',\varphi',x'),
\nonumber \\
\label{jkq-approx}
\end{eqnarray}
where
\begin{equation}
\int_{x',\bm{\Omega}'}=
\int_{-\infty}^{+\infty} dx' 
\oint \frac{d\bm{\Omega}'} {4 \pi},
\end{equation}
and $c_i, i=2,3,4,5,6$ are positive numbers (see appendix D of Paper III).
We recall that ${\rm d} \bm{\Omega}'=\sin \theta'\,d\theta'\,d\varphi'$, 
$\theta' \in [0, \pi]$ and $\varphi' \in [0, 2\pi]$. Here 
\begin{eqnarray}
&&\hat{R}(x,x',\bm{\Omega},\bm{\Omega}')=\nonumber \\ 
&&\hat{W}  \left[\hat{\alpha}r_{\rm II}(x,x',\bm{\Omega},\bm{\Omega}')
+\left(\hat{\beta}-\hat{\alpha}\right)
r_{\rm III}(x,x',\bm{\Omega},\bm{\Omega}')\right],
\label{redis-nonmag}
\end{eqnarray}
is the non-magnetic, polarized redistribution matrix.
\\

\noindent
Step 4:\, A Fourier expansion of the AD PRD
functions with respect to $\varphi'$ (instead of $\varphi$) gives
\begin{eqnarray}
&&r_{\rm II,III}(x,x',\bm{\Omega},\bm{\Omega'})=
\nonumber\\
&&\sum_{k'=0}^{k'=\infty}\,\, (2-\delta_{k'0}) 
e^{ik'\varphi'}\,\,
{\tilde{r}}_{\rm II,III}^{(k')}(x,x',\bm{\Omega},\theta'),
\label{2D-Fourier-series-r23-pf}
\end{eqnarray}
with the Fourier coefficients
\begin{eqnarray}
&&{\tilde{r}}^{(k)}(x,x',\bm{\Omega},\theta')
=\int_0^{2\pi}\frac{d\,\varphi'}{2\pi}\,
e^{-ik'\varphi'}\,\,
\nonumber\\
&&{r}_{\rm II,III}(x, x', \bm{\Omega}, \bm{\Omega'}).
\nonumber \\
\label{2D-f-r-coefficients}
\end{eqnarray}
Substituting Equation~(\ref{2D-Fourier-series-r23-pf}) in 
Equation~(\ref{jkq-approx}) we can show that
the components $J^{\rm 2,x}_1$ and $J^{\rm 2,y}_2$ do not vanish
irrespective of the symmetry of $I^0_0$ with respect to the
infinite spatial axis. In other words, to a first approximation, even if
we assume that $I^0_0$ is symmetric with respect to the infinite 
spatial axis (as in the AA PRD), the $\varphi'$-dependence of the 
AD PRD functions 
$r_{\rm II,III}$ is such that the integral over $\varphi'$ leads to non-zero
$J^{\rm 2,x}_1$ and $J^{\rm 2,y}_2$. This stems basically from the
coefficients with $k'\ne0$ in the expansion of the AD PRD functions. 
Following an induction proof as in Paper II, 
it follows that $J^{\rm 2,x}_1$ and $J^{\rm 2,y}_2$ are non-zero
in general because the symmetry breaks down in the first step itself.

It follows from Equation~(\ref{transform-1}), and from the above proof that
the Stokes $I$ parameter is not symmetric with respect to the infinite
spatial axis in a non-magnetic 2D media, in the AD PRD case, unlike the AA PRD
and CRD cases (see Appendix B of Paper II for the proof for the AA PRD).

%%%%%%%%%%%%%%%%%%%%%%%%%%%%%%%%%%%Table 1
\begin{table*}
\begin{center}
\caption{The dominant Fourier components contributing to each
of the 6 irreducible components of $\bm{\mathcal I}$ 
in a non-magnetic 2D medium, shown as cross symbols.}
\vspace{0.6cm}
\label{table_1}
\begin{tabular}{ccccccc}
\tableline\tableline
&$k=0$&$k=1$&$k=2$&$k=3$&$k=4$\\
$\tilde{\mathcal I}^{0(k)}_0$&x&-&-&-&-\\
$\tilde{\mathcal I}^{2(k)}_0$&x&-&-&-&-\\
${\mathcal Re}\left[\tilde{\mathcal I}^{2,\rm x(k)}_1\right]$
&x&x&-&-&-\\
${\mathcal Im}\left[\tilde{\mathcal I}^{2,\rm x(k)}_1\right]$
&-&-&-&-&-\\
${\mathcal Re}\left[\tilde{\mathcal I}^{2,\rm y(k)}_1\right]$
&x&-&-&-&-\\
${\mathcal Im}\left[\tilde{\mathcal I}^{2,\rm y(k)}_1\right]$
&-&x&-&-&-\\
${\mathcal Re}\left[\tilde{\mathcal I}^{2,\rm x(k)}_2\right]$
&x&-&x&-&-\\
${\mathcal Im}\left[\tilde{\mathcal I}^{2,\rm x(k)}_2\right]$
&-&-&-&-&-\\
${\mathcal Re}\left[\tilde{\mathcal I}^{2,\rm y(k)}_2\right]$
&-&-&-&-&-\\
${\mathcal Im}\left[\tilde{\mathcal I}^{2,\rm y(k)}_2\right]$
&-&-&x&-&-\\
\tableline
\tableline
\end{tabular}
\end{center}
\end{table*}
%%%%%%%%%%%%%%%%%%%%%%%%%%%%%%%%%%%%%Table 1

%%%%%%%%%%%%%%%%%%%%%%%%%%%%%%%%%%%%% Figure 2
%%%%%%%%%%%%%%%%%%%%%%%%%%%%%%%%%%%%%%%%%%%%%%%Figure 2
%%%%%%%%%%%%%%%%%%%%%%%%%%%%%%%%%%%%% 
%%%%%%%%%%%%%%%%%%%%%%%%%%%%%%%%%%%%%
%%%%%%%%%%%%%%%%%%%%%%%%%%%%%%%%%%%%% 
%%%%%%%%%%%%%%%%%%%%%%%%%%%%%%%%%%%%% 
%%%%%%%%%%%%%%%%%%%%%%%%%%%%%%%%%%%%% 
%%%%%%%%%%%%%%%%%%%%%%%%%%%%%%%%%%%%% 
%%%%%%%%%%%%%%%%%%%%%%%%%%%%%%%%%%%%% 
%%%%%%%%%%%%%%%%%%%%%%%%%%%%%%%%%%%%% 
%%%%%%%%%%%%%%%%%%%%%%%%%%%%%%%%%%%%% 
%%%%%%%%%%%%%%%%%%%%%%%%%%%%%%%%%%%%% 
%%%%%%%%%%%%%%%%%%%%%%%%%%%%%%%%%%%%% 
%%%%%%%%%%%%%%%%%%%%%%%%%%%%%%%%%%%%% 
%%%%%%%%%%%%%%%%%%%%%%%%%%%%%%%%%%%%% 
%%%%%%%%%%%%%%%%%%%%%%%%%%%%%%%%%%%%% 
%%%%%%%%%%%%%%%%%%%%%%%%%%%%%%%%%%%%% 
%%%%%%%%%%%%%%%%%%%%%%%%%%%%%%%%%%%%%

%%%%%%%%%%%%%%%%%%%%%%%%%%%%%%%%%%%%% 
%%%%%%%%%%%%%%%%%%%%%%%%%%%%%%%%%%%%% 
%%%%%%%%%%%%%%%%%%%%%%%%%%%%%%%%%%%%% 
%%%%%%%%%%%%%%%%%%%%%%%%%%%%%%%%%%%%% 

\end{document}